\documentclass[pre,preprint,amsmath,superscriptaddress,endfloats]{revtex4}

\usepackage{graphicx}
\usepackage{units}
\usepackage{amssymb}

\begin{document}

\title{Generating macroscopic chaos in a network of globally coupled phase oscillators}

\author{Paul So}
\email[Email address: ]{paso@gmu.edu}
\homepage[Web page: ]{http://complex.gmu.edu/~paso}
\affiliation{School of Physics, Astronomy, \& Computational Sciences, The Center for Neural Dynamics,
and The Krasnow Institute for Advanced Study, George Mason University, Fairfax
Virginia 22030, USA}

\author{Ernest Barreto}
\email[Email address: ]{ebarreto@gmu.edu}
\homepage[Web page: ]{http://complex.gmu.edu/~ernie}
\affiliation{School of Physics, Astronomy, \& Computational Sciences, The Center for Neural Dynamics,
and The Krasnow Institute for Advanced Study, George Mason University, Fairfax
Virginia 22030, USA}

\begin{abstract}
We consider an infinite network of globally-coupled phase oscillators in which the natural frequencies of the oscillators are drawn from a symmetric bimodal distribution. We demonstrate that macroscopic chaos can occur in this system when the coupling strength varies periodically in time. We identify period-doubling cascades to chaos, attractor crises, and horseshoe dynamics for the macroscopic mean field. Based on recent work that clarified the bifurcation structure of the static bimodal Kuramoto system, we qualitatively describe the mechanism for the generation of such complicated behavior in the time varying case.
\end{abstract}

\date{\today}


\maketitle

In nature and in many practical applications, it is not uncommon to observe the emergence of coherent macroscopic behavior in large populations of interacting rhythmic units despite noise and the presence of heterogeneity in the population. For systems of globally-coupled phase oscillators, collective synchrony and simple macroscopic oscillations have been extensively studied \cite{generalsynch,Kuramoto75,Kuramoto84,Crawford94,Strogatz00,Acebron05}.  Large networks of more complicated (e.g., chaotic) oscillators can also exhibit these behaviors \cite{collectivechaos}, but can also display a collective chaotic state \cite{chaoticmap}.  It is not clear, however, if heterogeneous networks constructed of \emph{simple} phase oscillators that are not independently capable of producing chaos can exhibit complex macroscopic behavior (such as chaos) in the thermodynamic limit of large system size. In the current work, we use a recently-developed mean-field analysis method and demonstrate that chaos can exist in the macroscopic mean field for a heterogeneous network of globally coupled phase oscillators with a bimodal frequency distribution and time-periodic coupling. We propose a qualitative mechanism for how this arises, and we identify period-doubling scenarios, attractor crises, and Smale horseshoe dynamics.

\section{Introduction}

A remarkably successful and analytically tractable model for describing the spontaneous onset of coherence in large populations of phase oscillators was introduced by Kuramoto in 1975 \cite{Kuramoto75,Kuramoto84}, and many subsequent extensions have been formulated \cite{Strogatz00,Acebron05}.  Typically, Kuramoto-like models share the following three fundamental characteristics: (i) The individual rhythmic units within the network are simple phase oscillators. When isolated, the temporal evolution of the oscillators is given by $\dot{\theta}_{i}=\omega_{i}, i=1,\cdots,N$, where $\omega_{i}$ is the intrinsic natural frequency of the \textit{i}-th oscillator, and the number of oscillators in the network (\textit{N}) is assumed to be large. (ii) The collection of natural frequencies is assumed to be distributed according to a time-invariant function $g(\omega)$. (iii) The coupling among oscillators is assumed to be all-to-all (i.e. global), so that each oscillator influences, and is influenced by, all others in the network.

The original Kuramoto model assumes a smooth unimodal distribution function $g(\omega)$ which is symmetric about a mean frequency $\omega_{0}$ and which monotonically approaches zero as $\lvert\omega-\omega_{0}\rvert\rightarrow\infty$.  When uncoupled, the network is incoherent, and each oscillator drifts in phase with respect to the others. But, when the coupling strength is sufficiently strong (and $N$ is sufficiently large), a coherent state emerges through a continuous phase transition at a critical coupling strength $K_*$. A macroscopic domain of phase-locked oscillators begins to form and coherence grows as the coupling strength $K$ continues to increase.  Traditionally, this phase transition is quantified by a complex order parameter $z$ (to be defined in Sec. (\ref{model_sec})) which describes the macroscopic mean field. This has magnitude zero in the incoherent state and becomes nonzero for $K>K_*$.

Previous efforts analyzing the emergence of coherence have mainly focused on the loss of stability of the incoherent state.  A recent breakthrough introduced by Ott and Antonsen \cite{OAreduction08,OAreduction09} has allowed us to move beyond this approach, and there have been many developments in understanding the Kuramoto system and its extensions \cite{Martens09,Marvel09,Pikovsky08,Paz06,PazoBimodal}.  The OA method identifies a low-dimensional invariant manifold to which the dynamics of the macroscopic mean field of a large ($N \rightarrow \infty$) heterogeneous network of globally-coupled phase oscillators is attracted. Remarkably, low-dimensional equations of motion for the mean-field behavior on this manifold can be derived. For example, the governing equation for the order parameter of the original Kuramoto problem (on the manifold) can be written as a single complex nonlinear ordinary differential equation.  Recent analyses of a number of Kuramoto-type models using this method have revealed a rich set of possible nonlinear dynamical states \cite{oaothers}: limit cycles, chimeras, quasi-periodic states, multistability, and standing waves, as well as various local and global bifurcation scenarios for the mean field, including saddle node, transcritical, Hopf, saddle-node infinite-period (SNIPER), and homoclinic bifurcations.

But can a network of simple phase oscillators result in more complicated dynamical behavior, such as chaos, in the infinite-$N$ limit? Kuramoto himself speculated that ``within the framework of the phase model\dots no chaotic dynamics at the collective level seems possible'' \cite{Kuramoto93}.
For globally-coupled complex Ginzburg-Landau-type oscillators, clustering and macroscopic chaos have been observed \cite{Matthews91,Kuramoto93,Kuramoto94,Takeuchi09}. 
However, it has been argued that this requires a degree of freedom in the oscillator amplitude \cite{Kuramoto93,Kuramoto94}, and thus the model diverges from the simple phase oscillator description. Chaotic behavior has also been reported in coupled map lattices \cite{chaoticmap}, but the individual elements of such systems are typically nonlinear maps that are separately capable of producing chaos when uncoupled. There is numerical evidence suggesting that chaos exists in \emph{finite} populations of a number of Kuramoto-type phase models \cite{Popovych05,Rapisarda09}, but this appears to disappear as $N \rightarrow \infty$. Chaos in the order parameter has also been reported for a resistively-loaded Josephson junction array \cite{Golomb92,Watanabe94}.  However, the analytic model in \cite{Watanabe94} corresponds to the non-generic case of a Kuramoto-like system with a \textit{homogeneous} (delta function) distribution of natural frequencies.  This is notable because it has been shown that the dynamics for the classical Kuramoto system \cite{OAreduction08} and the resistively loaded Josephson junction network \cite{Marvel09,OAreduction09} cannot posses a chaotic attractor in the infinite-$N$ limit in the more generic situation with a spread in the oscillator natural frequencies. 

The question is then: can attracting macroscopic chaos or other complicated dynamics exist in a \emph{heterogeneous} network of \emph{simple} phase oscillators in the thermodynamic limit ($N\rightarrow\infty$)? If so, what might the minimum requirements be? Here, we provide a partial answer by extending the bimodal Kuramoto model, which we analyzed previously \cite{Martens09}, to the case in which the global coupling parameter is a periodic function of time. We derive equations of motion for order parameters which are valid for $N\rightarrow\infty$ and long times, and we use these to firmly establish the existence of chaos in the macroscopic mean field by identifying period-doubling cascades to chaos, attractor crises, and Smale horseshoe dynamics.

\section{Formulation}
The bimodal Kuramoto system has been investigated by Kuramoto \cite{Kuramoto84}, Crawford \cite{Crawford94}, and others \cite{Bonilla92,Paz06,PazoBimodal}.
For the case of a bimodal natural frequency distribution $g(\omega)$ consisting of the sum of two offset but otherwise identical Cauchy-Lorentz distributions, a complete bifurcation diagram was recently reported \cite{Martens09}. (The same authors also found that when the sum of two offset Gaussians was used for $g(\omega)$, the resulting bifurcation diagram was qualitatively the same.)

Most related work assumes that the connectivity of the network of interest is static. However, in natural networks (i.e., physical,
biological, social, etc.), communication among the individual components is often time-dependent (see \cite{So08} and the references therein). The coupling strength, the type of coupling, and the connection topology may not necessarily
remain constant.  Here, we generalize the bimodal Kuramoto system to include the situation in which the global coupling parameter varies periodically in time.  

\subsection{Our Model}
\label{model_sec}

We consider the following system:
\begin{equation}
\dot{\theta_i}=\omega_i + \frac{K(t)}{N} \sum_{j=1}^{N} \sin(\theta_j - \theta_i)
\label{ourmodel_eq}
\end{equation}
for $i=1, 2, \dots, N$, in which the coupling strength is assumed to be a periodic function given by
\begin{equation}
K(t)=K_0+A\sin\left(\frac{2\pi}{\tau}t\right).
\label{kvary_eq}
\end{equation}
The strength and the period of the periodic variation are given by $A$ and $\tau$ respectively.  These two parameters will serve as the bifurcation parameters in the current study.

\begin{figure}
\begin{center}
\scalebox{0.8}{\includegraphics{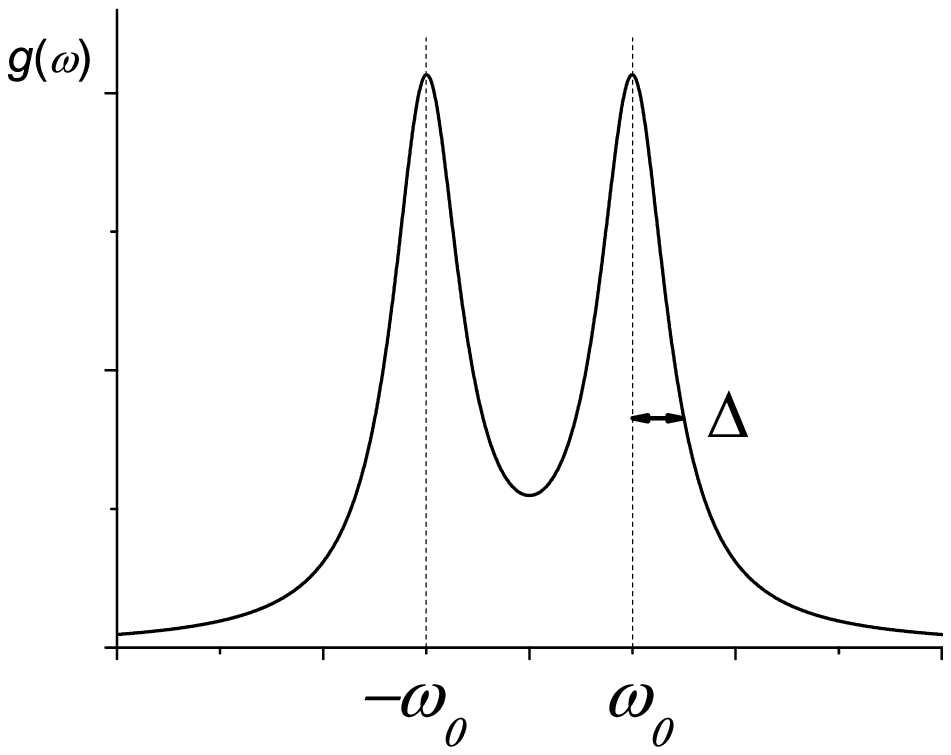}}
\caption{Bimodal distribution of natural frequencies, $g(\omega)$, as a sum of two Lorentzians.}
\label{bimodal_fig}
\end{center}
\end{figure}

The natural frequency $\omega_i$ for each oscillator is randomly drawn from a normalized bimodal
distribution function given by the sum of two Cauchy-Lorentz distributions,
\begin{equation}
g(\omega)=\frac{\Delta}{2\pi}\left( \frac{1}{(\omega-\omega_0)^2+\Delta^2}+\frac{1}{(\omega+\omega_0)^2+\Delta^2}\right),
\label{freqdist_eq}
\end{equation}
where $\Delta$ characterizes the half-width of the individual peaks and $\pm\omega_0$ are their center frequencies (see Fig. \ref{bimodal_fig}).  Without loss of generality, we choose $g(\omega)$ to be symmetric about $\omega=0$, as one can always pick an appropriate rotating frame in which this holds.  However, for $g(\omega)$ to be effectively bimodal, one must choose $\omega_0>\Delta/\sqrt{3}$ so that the central dip of the bimodal distribution is convex. A more general study of the Kuramoto system with an asymmetric bimodal distribution will be reported elsewhere \cite{Cotton10}, and a case in which a bimodal form of $g(\omega)$ that cannot be written as the sum of two even unimodal distributions was analyzed in \cite{PazoBimodal}.

\subsection{OA reduction}
In the infinite-$N$ limit, we can describe the phase oscillators within our network at a particular time $t$ by a continuous distribution function $F(\theta,\omega,t)$, such that $F(\theta,\omega,t)d\theta d\omega$ gives the fraction of phase oscillators with 
phases between $\theta$ and $\theta+d\theta$ and with natural frequencies between $\omega$ and $\omega+d\omega$.  
Since the number of oscillators is assumed to be conserved at all times, we have
\[
\int_{-\infty}^{\infty}\int_{0}^{2\pi}F(\theta,\omega,t)\,d\theta d\omega = 1,
\]
and since $g(\omega)$ is assumed to be constant in time, the marginal distribution density
\[
\int_{0}^{2\pi}F(\theta,\omega,t)\,d\theta = g(\omega)
\]
is as well.  Many authors have analyzed the evolution of Kuramoto-type systems in terms of this distribution function (e.g., \cite{Kuramoto84,Strogatz00,Crawford94,OAreduction08,OAreduction09}).  In particular, the time evolution of $F$ must satisfy the continuity equation,
\begin{equation}
\label{continuity_eq}
\frac{\partial F}{\partial t} + \frac{\partial}{\partial \theta}\left(F v_\theta\right)=0,
\end{equation}
where the phase velocity $v_\theta$ is given by the continuum version of Eq.~\ref{ourmodel_eq},
\begin{equation}
\label{velfield1_eq}
v_\theta(\theta,\omega,t)=\omega+K(t)\int_{0}^{2\pi}F(\theta,\omega,t)\sin(\theta'-\theta)d\theta.
\end{equation}
The macroscopic mean field is described by a complex order parameter originally introduced by Kuramoto \cite{Kuramoto84},
\begin{equation}
\label{orderpar_eq}
z(t)=\rho e^{i\psi}=\int_{-\infty}^{\infty}\int_{0}^{2\pi}F(\theta,\omega,t)e^{i\theta}\,d\theta d\omega.
\end{equation}
Geometrically, the order parameter describes the centroid of all the phasors $e^{i\theta}$ within the network.  When the network is incoherent, $z(t)$ is zero; it becomes nonzero when coherence emerges.  Our goal is to describe the potentially complicated dynamical states for this macroscopic variable when we vary the system parameters $A$ and $\tau$. 

The phase velocity $v_\theta$ can be re-written in terms of $z(t)$ as follows:
\begin{equation}
\label{velfield_eq}
v_\theta(\theta,\omega,t)=G+\frac{1}{2i}\left[H(t)e^{-i\theta}-H^*(t)e^{i\theta}\right],
\end{equation}
where $H(t)=K(t)z(t)$ and $G=\omega$.  Note that due to the global coupling, $H(t)$ does not explicitly depend on the individual phase $\theta$ except through the mean field $z(t)$. In fact, one can take Eq. (\ref{velfield_eq}) as the starting point for a generalization to a larger class of Kuramoto-type systems for which the OA reduction method is applicable. The method applies as long as the network vector field can be put into the single harmonic form given by Eq. \ref{velfield_eq}, where neither $H$ nor $G$ depend on $\theta$ explicitly.

Eqs.~(\ref{continuity_eq}), (\ref{orderpar_eq}), and (\ref{velfield_eq}) form the governing equations for the distribution function $F(\theta,\omega,t)$.  Following Ref.~\cite{OAreduction08}, we expand $F(\theta,\omega,t)$ in a Fourier series according to the ansatz
\begin{equation}
\label{ansatz_eq}
F(\theta,\omega,t)=\frac{g(\omega)}{2\pi}\left[1+\sum_{n=1}^{\infty}\alpha^{n}(\omega,t)e^{in\theta}+\sum_{n=1}^{\infty}\alpha^{*n}(\omega,t)e^{-in\theta}\right],
\end{equation}  
where $\alpha(\omega,t)$ is a yet-to-be-determined function independent of $\theta$ and which has modulus less than one (so as to guarantee convergence of the Fourier series).  It is important to note that this ansatz defines a submanifold within the infinite-dimensional space of all possible distribution functions $F$.  By directly substituting the ansatz, Eq. (\ref{ansatz_eq}), into the continuity equation, Eq.~(\ref{continuity_eq}), one obtains
\begin{equation}
\label{condition_eq}
\left(\frac{\partial\alpha}{\partial t}+i\omega\alpha+\frac{1}{2}\left[H\alpha^2-H^*\right]\right)\sum_{n=1}^{\infty}n\alpha^{n-1}e^{in\theta}+c.c.=0,
\end{equation}
where c.c. denotes the complex conjugate of the expression on the left-hand-side of the equation. Then, since $\sum_{n=1}^{\infty}n\alpha^{n-1}e^{in\theta}=\frac{e^{i\theta}}{(1-\alpha e^{i\theta})^2}\neq0$, 
the factor within the parentheses must vanish identically if the ansatz given by Eq.~(\ref{ansatz_eq}) is to be a valid solution to the continuity equation.  This then gives
\begin{equation}
\label{solution_eq}
\frac{\partial \alpha}{\partial t}+i\omega\alpha+\frac{1}{2}\left[H\alpha^2-H^*\right]=0,
\end{equation}
which holds for each value of $\omega$.  Note that $H$ will in general depend on the macroscopic mean field $z(t)$, which can be written
\begin{equation}
\label{solution_mean_eq}
z(t)=\int_{-\infty}^{\infty}\int_{0}^{2\pi}F(\theta,\omega,t)e^{i\theta}\,d\theta d\omega=\int_{-\infty}^{\infty}\alpha^*(\omega,t)g(\omega)d\omega.
\end{equation}
Therefore, the time evolution of $\alpha(\omega,t)$, and thus the distribution function $F(\theta,\omega,t)$, can be obtained by integrating the integro-differential equation given by Eqs.~(\ref{solution_eq}), (\ref{solution_mean_eq}) and the mean field function $H(z)$ \cite{noteattract}.  Note that although the use of the ansatz has effectively collapsed the time evolution of the infinite number of Fourier modes in $\theta$ into one evolution equation for $\alpha$, Eqs. (\ref{solution_eq}) and (\ref{solution_mean_eq}) remain an infinite-dimensional system since we have a continuous set of natural frequencies $\omega$.  As we will show in our specific examples below, Eqs.~(\ref{solution_eq}) and (\ref{solution_mean_eq}) reduce further to only a small number of ordinary differential equations.

\subsection{No Chaos in the Classic Kuramoto Model}

The original Kuramoto model can be put into the form of Eq.~(\ref{velfield_eq}) if we define $H(t)=Kz(t)$, $G=\omega$,  and let $\omega$ be chosen randomly from a unimodal Cauchy-Lorentz distribution.
We then have
\[
v_\theta(\theta,\omega,t)=\omega+\frac{K}{2i}\left[ze^{-i\theta}-z^*e^{i\theta}\right].
\]
Then, following the formalism described above, the amplitude $\alpha(\omega,t)$ evolves in time according to
\[
\frac{\partial \alpha}{\partial t}+i\omega\alpha+\frac{K}{2}\left[z\alpha^2-z^*\right]=0,
\]
with the mean field $z(t)$ being given by Eq.~(\ref{solution_mean_eq}).
Following Ref.~\cite{OAreduction08}, we further assume that $\alpha(\omega,t)$ can be analytically continued into the lower complex $\omega$-plane and that the initial conditions satisfy (i) $|\alpha(\omega,0)|\leq1$, and (ii) $|\alpha(\omega,0)|\rightarrow0$ as $\Im(\omega)\rightarrow\-\infty$.  Then, using a semi-circular contour in the lower complex $\omega$-plane to perform the integral in Eq. (\ref{solution_mean_eq}), and taking the radius of the contour to infinity, we can express $z(t)$ in terms of the residue of the contour integral, i.e., $z(t)=\alpha^*(\omega_0-i\Delta,t)$.
Therefore, at large time, the macroscopic mean field for the classical Kuramoto system evolves according to a single complex ordinary differential equation:
\begin{equation}
\label{classic_eq}
\frac{dz}{dt}=-(\Delta+i\omega_0)z+\frac{K}{2}(z-z^*z^2).
\end{equation}

Since this is a two-dimensional system, no complicated behavior such as chaos is possible for the classical Kuramoto system with a heterogeneous distribution of natural frequencies in the thermodynamic limit. This result does not contradict the numerical results obtained in Refs.~\cite{Popovych05,Rapisarda09}, since those examples considered finite-size networks.  In fact, numerical evidence in Ref.~\cite{Popovych05} indicated that the largest Lyapunov exponent in the finite-size network tends toward zero as the system size increases.

\subsection{Bimodal Kuramoto Model with Time-varying Coupling}

The same functions $H(t)$ and $G$ used above apply to the bimodal case with time-dependent coupling. The difference is the form of the distribution function $g(\omega)$. Following the same procedure outlined above leads to
\begin{equation}
\label{bi_pdf_eq}
\frac{\partial \alpha}{\partial t}+i\omega\alpha+\frac{K(t)}{2}\left[z\alpha^2-z^*\right]=0,
\end{equation}
where the coupling $K(t)$ is now time-dependent according to Eq.~(\ref{kvary_eq}) and the mean field $z(t)$ is given by Eq.~(\ref{solution_mean_eq}).
Since explicit time dependence appears only in $H(t)=K(t)z(t)=(K_0+A \sin(2\pi t/\tau)z(t)$, the mathematical analysis is similar to the time-independent case reported in \cite{Martens09}. We briefly summarize the procedure and the relevant results.

The bimodal Lorentzian distribution has two simple poles in the lower $\omega$-complex plane, $\omega=\pm\omega_0-i\Delta$, and direct integration of
Eq.~(\ref{solution_mean_eq}) gives $z(t)=\frac{z_1(t)+z_2(t)}{2}$ with $z_{1,2}(t)=\alpha^*(\pm\omega_0-i\Delta,t)$. These two sub-order parameters represent the two populations of phase oscillators whose natural frequencies cluster around $\omega=+\omega_0$ and $\omega=-\omega_0$, respectively \cite{Barreto07,Martens09}.
Substituting this expression for $z(t)$ into the differential equation for $\alpha$, we arrive at the following pair of nonautonomous equations for $z_1(t)$ and $z_2(t)$:
\begin{equation}
\label{temp2_eq}
\begin{array}{ccl}
\frac{dz_1}{dt} &=& -(\Delta+i\omega_0)z_1 + \frac{K(t)}{4}\left[z_{1}+z_{2}-(z^*_1+z^*_2)z_1^2)\right]\\
\frac{dz_2}{dt} &=& -(\Delta-i\omega_0)z_2 + \frac{K(t)}{4}\left[z_{1}+z_{2}-(z^*_1-z^*_2)z_2^2)\right].
\end{array}
\end{equation}
We expect the asymptotic behavior of these two complex sub-order parameters to be symmetric except for a relative phase difference $\psi$ such that $\frac{z_2}{z_1}=e^{i\psi}$ and $|z_2|=|z_1|=\rho$ (for details, see \cite{OAreduction09} and \cite{Martens09}). Substituting this symmetry condition into Eq.~({\ref{temp2_eq}), we obtain equations that describe the long-time macroscopic mean-field behavior for the time-varying bimodal Kuramoto system:
\begin{equation}
\label{bimodal_eq}
\begin{array}{ccl}
\dot{\rho} &=& -\Delta\rho+\frac{K(t)}{4}\rho(1-\rho^2)(1+\cos\psi)\\
\dot{\psi} &=& 2\omega_0-\frac{K(t)}{2}(1+\rho^2)\sin\psi.
\end{array}
\end{equation}
Note that the OA reduction of the time-static network in \cite{Martens09} has the same mathematical form as this, except that here, $K(t)$ is a function of time. The full bifurcation structure of these equations with $K(t)=constant$ was analyzed in \cite{Martens09}, and the main features are shown in Fig.~\ref{phase_diagram_fig}. For convenience, we refer to this system as the ``static" system, and we write ``time-varying system" to describe the case when $K(t)$ varies in time according to Eq.~(\ref{kvary_eq}).

\begin{figure}
\begin{center}
\scalebox{1.0}{\includegraphics{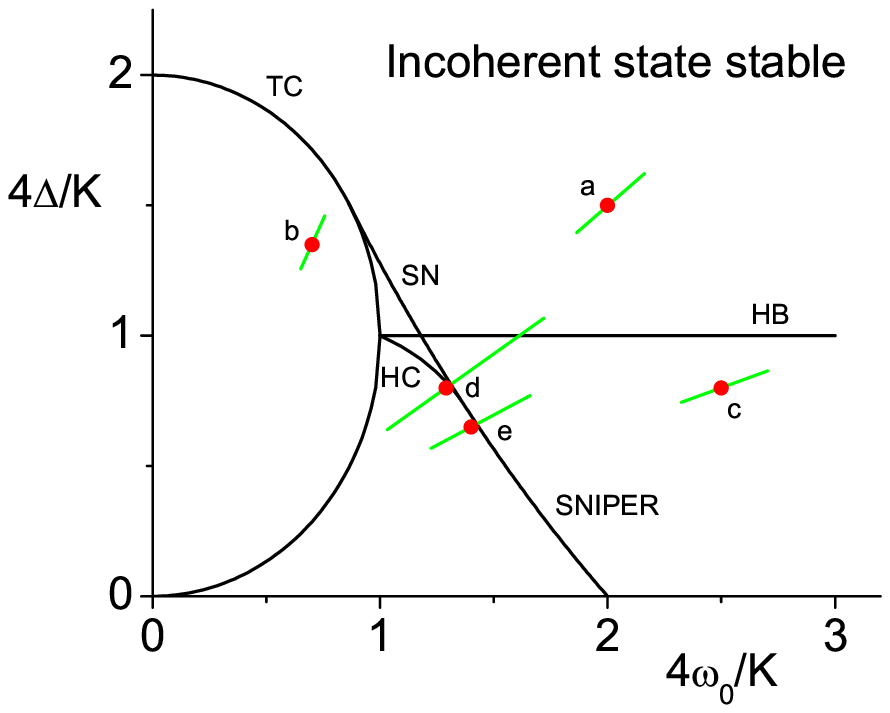}}
\caption{(Color online) Phase diagram for the static bimodal Kuramoto network with coupling $K=K_0$. The incoherent state is stable in the upper region, and black curves denote bifurcations that lead to coherent collective states (TC=transcritical, HB=Hopf, SN=saddle-node, HC=homoclinic, SNIPER=saddle-node-infinite-period). In the time-varying system, $K(t)=K_0+A\sin(2\pi t/\tau)$, and the parameters sweep along the short diagonal green lines. The lettered red points indicate cases of interest for which $(\omega_0,\Delta)$ are as follows: (a) $(2.0,1.5)$; (b) $(0.7,1.35)$; (c) $(2.5,0.8)$; (d) $(1.29,0.8)$; (e) $(1.4,0.65)$. We set $K_0=4$ and we have $A=0.3$ for cases (a)-(c) and $A=1$ for case (d).}
\label{phase_diagram_fig}
\end{center}
\end{figure}

An important result of Ref.~\cite{Martens09}, which analyzed the static version of Eqs.~(\ref{bimodal_eq}), is that two particular combinations of parameters, $4\omega_0/K$ and $4\Delta/K$, reveal the full dynamical structure of the system. Figure \ref{phase_diagram_fig} shows the parameter space described by these expressions. The incoherent state is stable for parameter values at the top of this figure. Its stability can be lost either via a transcritical bifurcation (TC, semicircular boundary) or by a Hopf bifurcation (HB, half-line). A saddle-node (SN) bifurcation occurs in the vicinity of where the TC and the HB line meet (but away from the incoherent state in state space), and thus there is an approximately triangular region of multistability in which an attracting partially-synchronized state coexists with the attracting incoherent state. The multistable region extends below the Hopf curve to the homoclinic (HC) bifurcation curve until the SN curve joins it and becomes a SNIPER curve.

Eqs.~(\ref{bimodal_eq}) describe the macroscopic mean field of the full network (Eqs.~(\ref{ourmodel_eq}-\ref{kvary_eq})) with $N \rightarrow \infty$ once the dynamics has converged onto the manifold defined by the ansatz (Eq.~\ref{ansatz_eq}).
Ref. \cite{OAreduction09} showed that convergence onto this ansatz manifold occurs as long as there is heterogenity in the natural frequencies of the oscillators. We note that Eqs.~(\ref{bimodal_eq}) describe the dynamics on the ansatz manifold for both the static and the time-varying system, since the the time-dependent coupling that we introduce in Eq.~(\ref{kvary_eq}) carries through in the reduction procedure. In particular, $K(t)$ does not produce excursions transverse to the ansatz manifold. However, under circumstances to be examined below, it produces complex trajectories 
\emph{within} the ansatz manifold. Thus, we expect these complex trajectories to be observable in the full network.

\section{Macroscopic Dynamics from a Time-Varying Bimodal Network}

Below, we use Eqs.~(\ref{bimodal_eq}) to analyze the asymptotic dynamics of the macroscopic mean field $(\rho, \psi)$ with the time-dependent coupling strength given by Eq.~(\ref{kvary_eq}). 
To understand these dynamics, it is useful to consider the relationship between the attractors of the static network and those of the time-varying system.
Recall that in the latter, the coupling strength varies periodically according to $K(t)=K_0+A\sin\left(2\pi t/\tau \right)$.
Thus, as $K(t)$ varies (with fixed values of $\omega_0$ and $\Delta$), the parameters in Figure \ref{phase_diagram_fig} vary and sweep through one of the regions indicated by the diagonal green lines. The attractors of the static system become ``moving targets" in the time-varying system.
We will refer to these ``moving targets" as pseudo-static attractors.
The resulting behavior of the time-varying system depends on the amplitude $A$ and frequency $1/\tau$ of the time-varying coupling. 
We consider the five cases labeled (a) through (e) marked by the red points in Figure \ref{phase_diagram_fig}, which indicate the static parameter values $(\omega_0, \Delta)$ for the average value of $K(t)$ in each case. We set $K_0=4$ throughout. See the figure caption for the remaining parameter values.

\subsection{The Incoherent State and Periodic and Quasiperiodic Solutions}

In cases (a)-(c), we set $\tau=5$ and $A=0.3$. Note that the amplitude $A$ is sufficiently weak such that the time-dependent system parameters remain fully within each respective region of stability relative to the static network.

For case (a), both the static system and the time-dependent system possess a fixed attracting equilibrium at $\rho=0$ corresponding to the incoherent state. Although the eigenvalues of this equilibrium oscillate in the time-varying system, it remains an attracting state, and hence the long-time behavior is not affected by the time variation of the coupling.

At the parameter values for case (b), the static system is attracted to an equilibrium that corresponds to a partially-synchronized state with $\rho > 0$. In contrast to case (a), this equilibrium is destroyed by the time variation of the coupling. In its place, the time-varying system exhibits a periodic orbit
with period $\tau$. This arises as the time-varying system's trajectory follows the ``moving target'', the pseudo-static equilibrium.
The link between equilibria of the static network and periodic orbits of the time-varying network (away from bifurcations) can be made rigorous by applying standard averaging methods (see, e.g., \cite{Gluckenheimer_holmes} or \cite{Sander_verhulst}).

Upon crossing the Hopf bifurcation line from above, the incoherent state of the static network loses stability, and an attracting periodic orbit emerges. This periodic orbit is the only attractor in case (c) for the static netowrk. For the time-varying system, the frequency of variation $1/\tau$ is typically not commensurate with the (effective) intrinsic frequency of oscillation of this periodic orbit. Thus, the asymptotic state for the time-varying network in this case is generally a quasiperiodic orbit on a torus \cite{trapping}. We also observed frequency locking behavior in this case as the amplitude $A$ was varied (not shown) \cite{quasiperiodic_future}.

We numerically confirmed that the time-dependent reduced system, Eq.~(\ref{bimodal_eq}), and the full system, Eq.~(\ref{ourmodel_eq}), both exhibit these local dynamical features. In the latter case we used an ensemble of $500,000$ oscillators. As expected, the weakly time-varying networks appropriately ``follow'' the expected behavior of the static $K_0$ networks. Fig.~\ref{sample_state_fig} compares the asymptotic behavior of the macroscopic mean field for the reduced and the full systems in cases (a)-(c).
\begin{figure}
\begin{center}
\scalebox{0.7}{\includegraphics{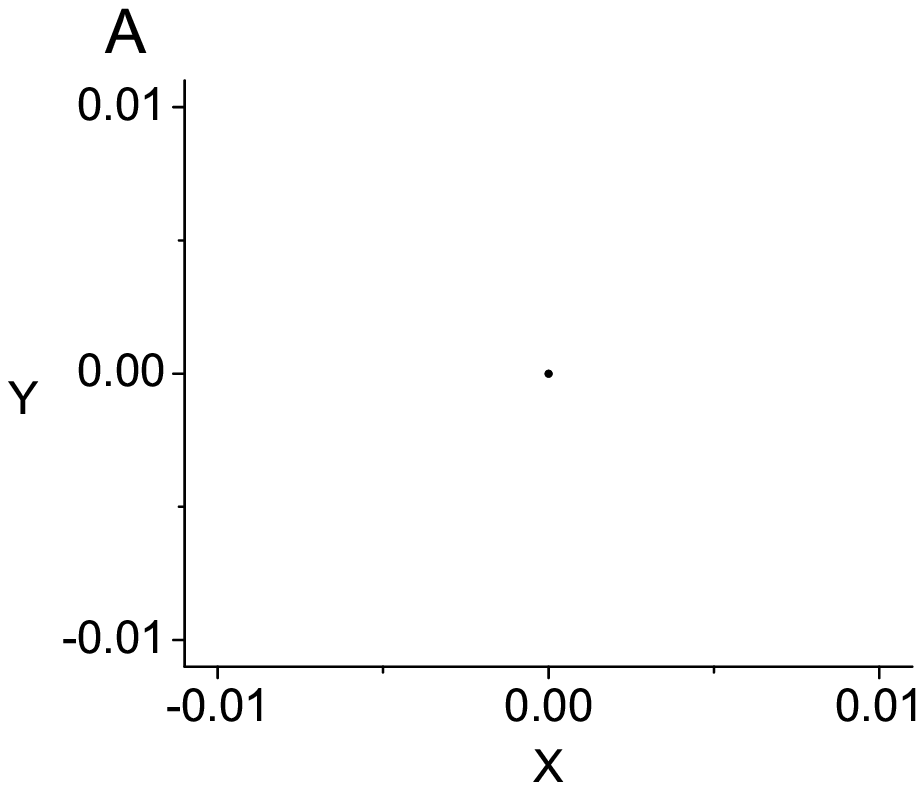}}
\scalebox{0.7}{\includegraphics{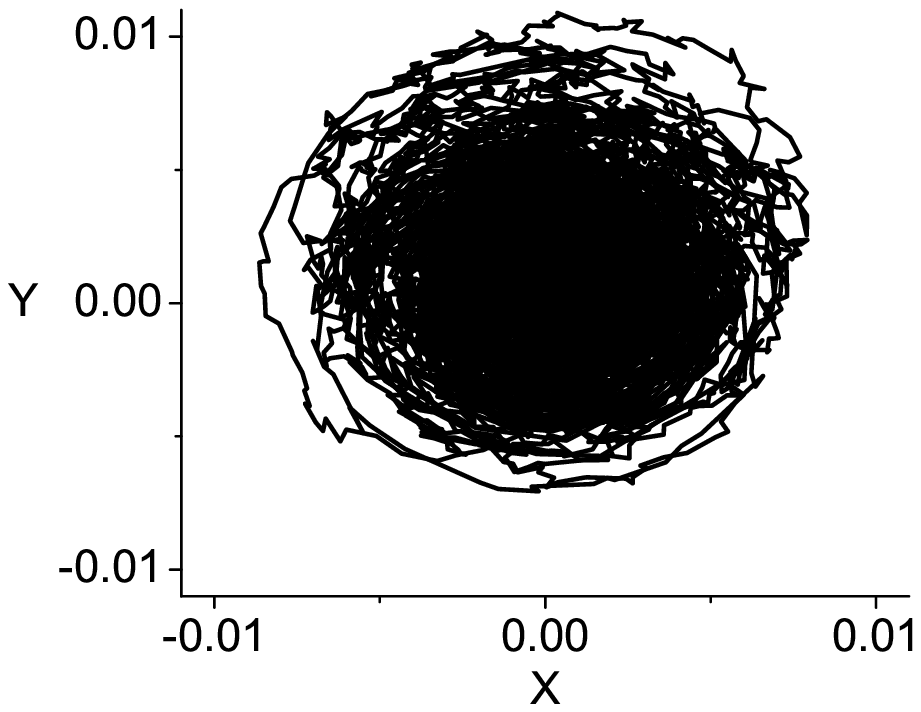}}
\end{center}
\begin{center}
\scalebox{0.7}{\includegraphics{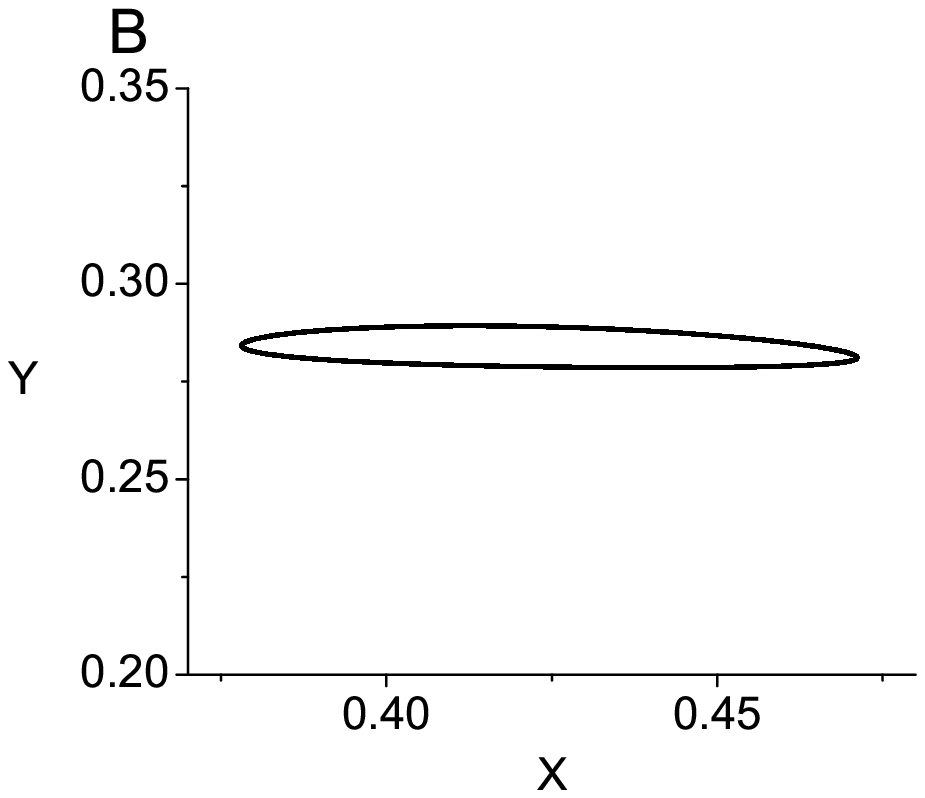}}
\scalebox{0.7}{\includegraphics{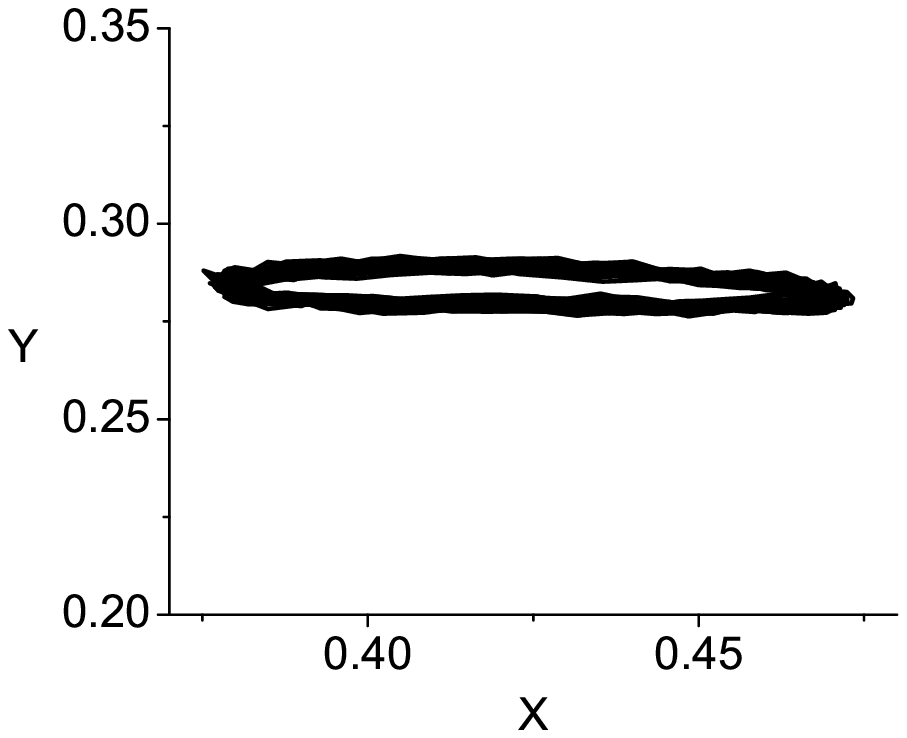}}

\end{center}
\begin{center}
\scalebox{0.7}{\includegraphics{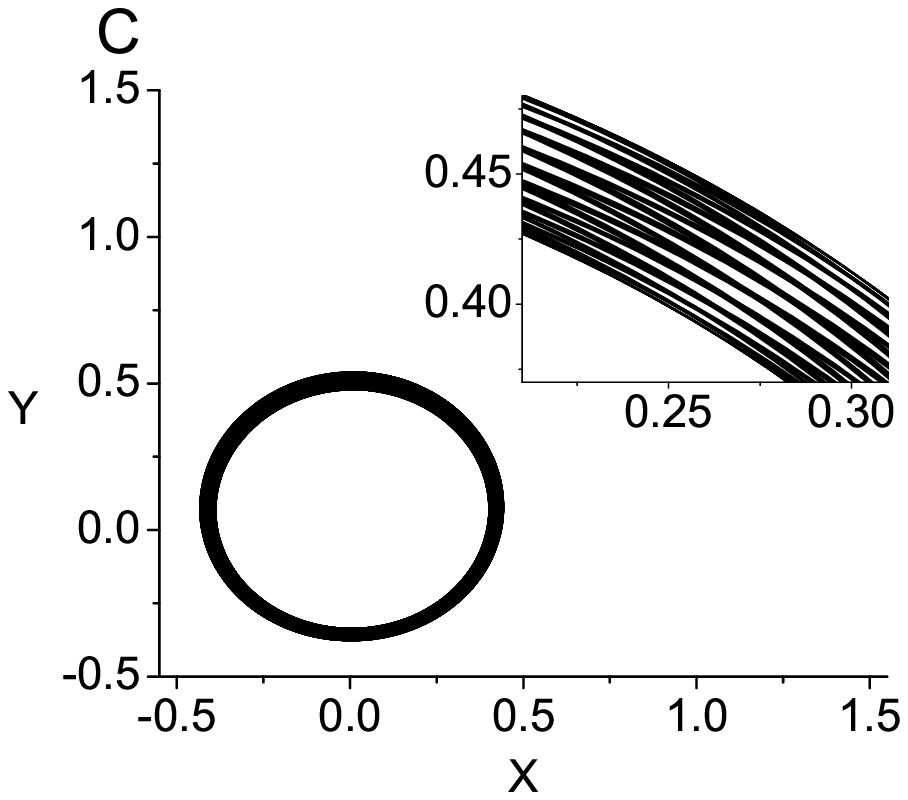}}
\scalebox{0.7}{\includegraphics{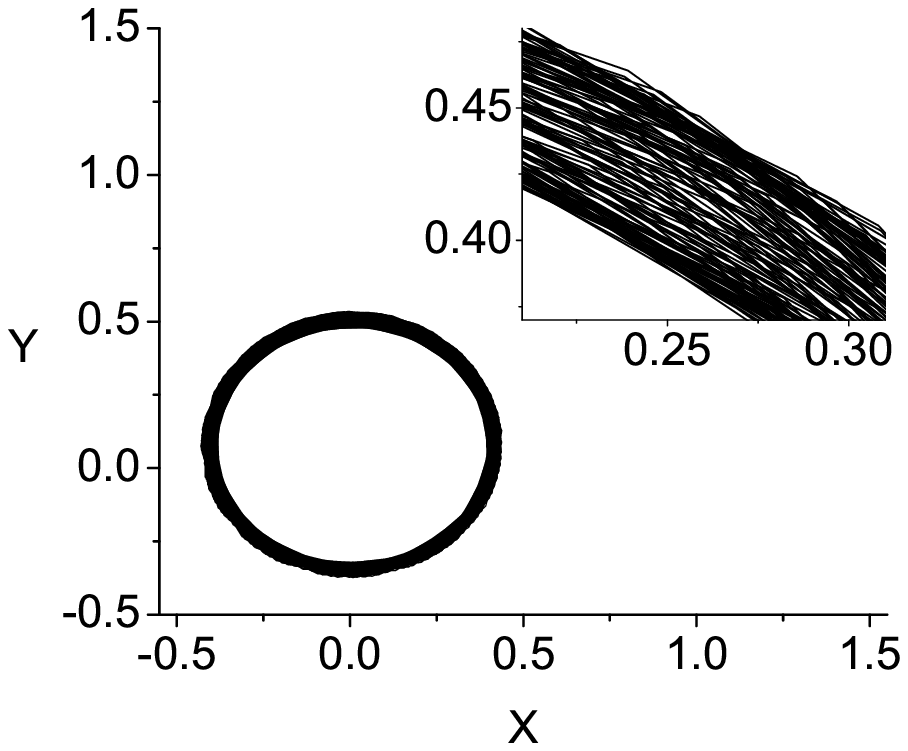}}

\end{center}
\begin{center}
\caption{Behavior of the reduced (left) and full (right) time-varying equations for cases (a)-(c) with $K_0=4$, $\tau=5$, and $A=0.3$.
A: Case (a), the persistence of the incoherent state. The fluctuations in the full system are due to finite-system-size effects. B: Case (b), a stable fixed point in the static system becomes a limit cycle (libration) in the weakly time-varying system. C: case (c), a limit cycle in the static system becomes a quasi-periodic state in the weakly time-varying system.}
\label{sample_state_fig}
\end{center}
\end{figure}
In case (a), both systems converge to the incoherent state, but the full system exhibits fluctuations on the order of $1/\sqrt{N}$ due to finite-system-size effects. The same effects can be seen in case (b), in which the full-system limit cycle is ``thickened" by a similar amount. In case (c), the apparent thickness of the ring-like attractor is much larger and reflects deterministic quasiperiodic dynamics as predicted by the reduced analysis (see inset).

These results are not surprising in light of the averaging theorem \cite{Verhulst96} and the results of Ref.~\cite{OAreduction09}. But, we emphasize that the dynamics induced by the periodic variation in $K(t)$ remain within the ansatz manifold, and we have confirmed that the behavior of the time-dependent reduced equations is evident in the full system. Of greater interest is the behavior of the time-varying system when the parameters sweep across bifurcations. We examine this in the next section.

\subsection{Period Doubling Cascades, Chaos, and a Crisis}

Case (d) is different from the previous cases in that the parameters sweep across multiple bifurcations as well as a small region of multistability. 
We note that as $K=K_0+A \sin(2 \pi t/\tau)$ varies in the range $[K_0-A,K_0+A]$ with $K_0=4$ and $A=1$, the parameters effectively sweep back and forth along the green line across this region for case (d) in Fig. \ref{phase_diagram_fig}. To clarify the situation, we show in Figure \ref{sweepbifdiag} (A)
\begin{figure}
\begin{center}
\scalebox{0.8}{\includegraphics{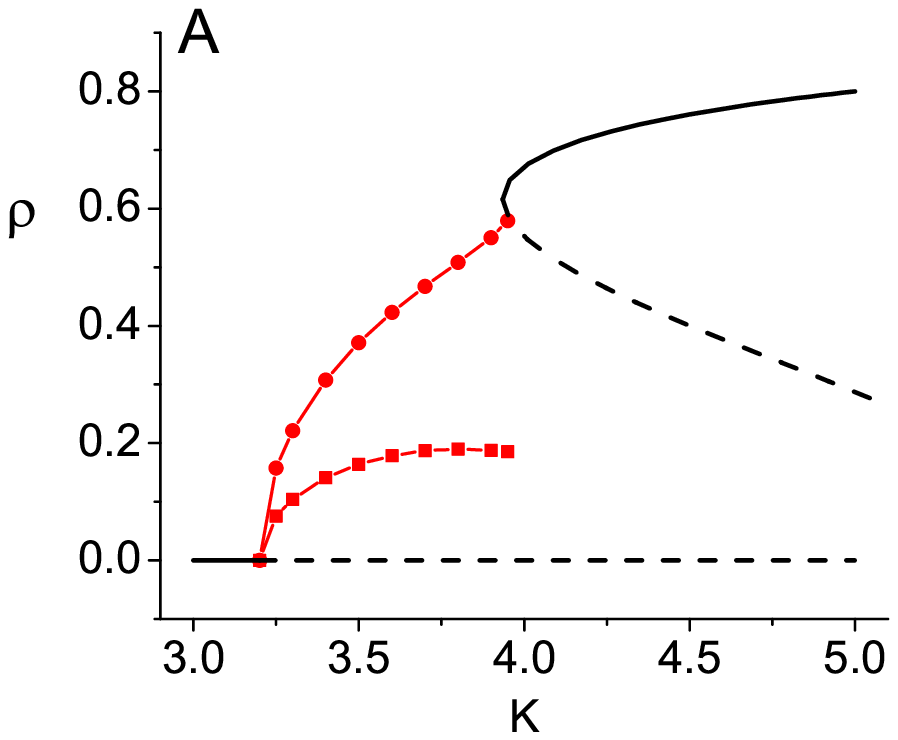}\includegraphics{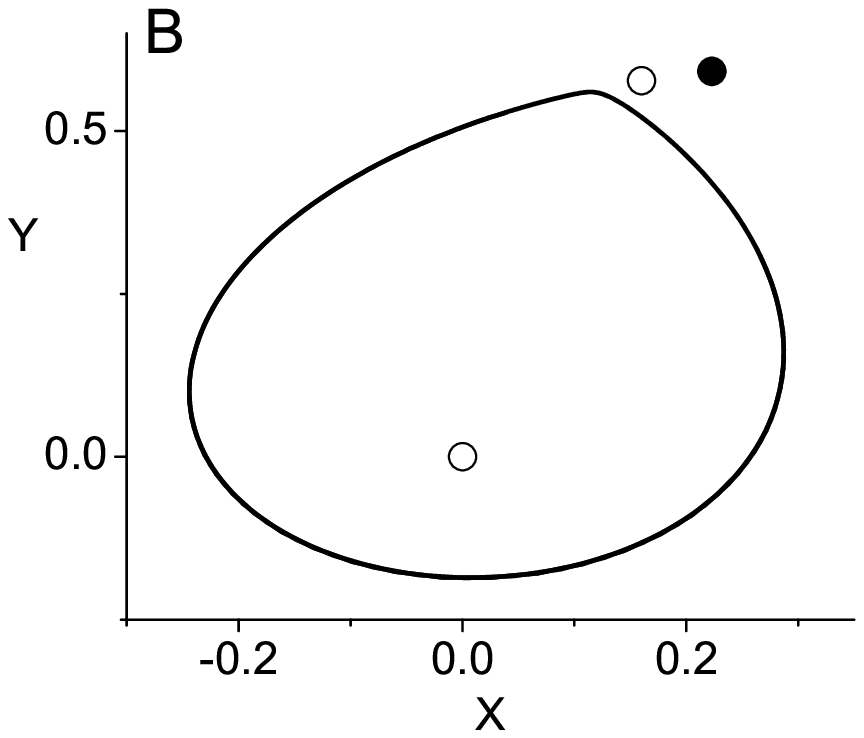}}
\caption{(Color online) A: Bifurcation diagram showing the static asymptotic structures past which we sweep. Solid lines are stable equilibria; dashed lines are unstable equilibria; lines with symbols represent the maxima (circles) and minima (squares) of limit cycles. B: State space diagram showing the limit cycle and the equilibria (circles; solid for stable and open for unstable) for K=3.94, when these coexist. $X=\rho \cos\psi$, $Y=\rho \sin\psi$, and in both panels, $\Delta=0.8$ and $\omega_0=1.29$.}
\label{sweepbifdiag}
\end{center}
\end{figure}
a one-dimensional bifurcation diagram, calculated using the static system with $K=K_0\pm A$, in which the magnitude of the asymptotic macroscopic mean-field $\rho$ is plotted versus $K$. The left panel shows the static asymptotic structures at each fixed parameter value $K$. The incoherent state is present throughout, and is attracting for $K<3.2$. At $K=3.2$, the Hopf bifurcation is encountered, and a periodic orbit (whose minimum and maximum $\rho$ values are plotted) emerges. At $K=3.935$, a saddle-node bifurcation creates two equilibria, and the unstable one collides with and destroys the periodic orbit in a homoclinic bifurcation at $K=3.953$. For larger values of $K$, only the equilibria exist. Fig.~\ref{sweepbifdiag}(B) shows all three static structures in state space for $K=3.94$, when they co-exist ($X=\rho \cos\psi$ and $Y=\rho \sin\psi$ are plotted).

The dynamics of the time-varying system can be understood in terms of transitory attraction to the pseudo-static attractors as $K(t)$ varies. For slow sweeping ($\tau$ small), there is ample time for the state of the time-varying system to approach the pseudo-static attractors as they drift in time. Thus the system either stays close to the pseudo-static attracting equilibrium or approximates the pseudo-static periodic orbit, depending on which is present at any given time. For extremely slow sweeping, this results in behavior similar to a bursting neuron, in which periods of quiescence alternate with ``bursts" of near-periodic excursions in phase space (see the movie in Fig.~(\ref{movie_slow}) in the Appendix). For very fast sweeping, the pseudo-static attractors come and go too quickly to affect the dynamics, and the time-varying system's behavior approaches that of the time-averaged system, i.e., the static system with $K=\left< K(t) \right>=K_0$ (see the movies in Figs.~(\ref{movie_fast}) and (\ref{movie_vfs}), Appendix) \cite{fastswitching,So08}. For intermediate sweeping frequencies, however, transitory attraction to moving pseudo-static attractors can dominate the dynamics of the time-varying system and complicated dynamics can arise, as we show below.

Using the reduced system, Eq.~(\ref{bimodal_eq}), we first investigate the time-varying network behavior by fixing $\tau=5$ (corresponding to an intermediate sweeping frequency) and gradually increasing the amplitude $A$ of the periodic coupling strength variation from zero. Accordingly, we sweep across the pseudo-static attractors shown in Figure \ref{sweepbifdiag}(A), from $-A$ to $A$, for increasing values of $A$. We investigate changes in behavior associated with $\tau$ later in this section.

For $A=0$, the time-varying system reduces to the static system with $(\omega_0, \Delta, K_0)=(1.29,0.8,4)$. In this case, the static system is attracted to an equilibrium with $\rho \neq 0$. For small values of $A>0$, the macroscopic mean-field exhibits a small limit cycle (libration) circling near the static equilibrium (not shown, but similar to Figs.~\ref{sample_state_fig} B). As $A$ increases, a saddle-node bifurcation occurs at $A=0.441$, creating a new stable limit cycle with full rotation in phase space. This limit cycle, shown in Fig.~\ref{period_double_fig}(A) for $A=0.55$, is 1:1 phase-locked to the drive period $\tau$. As $A$ increases further, a period-doubling occurs at $A=0.558$ (Fig.~\ref{period_double_fig}(B)), and then again at $A=0.600$ (Fig.~\ref{period_double_fig}(C)).  This period-doubling cascade continues, and by following the sequence through period sixteen and using the Feigenbaum constant \cite{Feigenbaum,Ottbook}, we estimate the infinite-period accumulation point to be at $A_\infty=0.615$. Beyond this, chaos is found. Fig.~\ref{period_double_fig}(D) shows the chaotic attractor obtained for $A=0.65$ (see the movie in Fig.~(\ref{movie_chaos}), Appendix).

\begin{figure}
\begin{center}
\scalebox{0.7}{\includegraphics{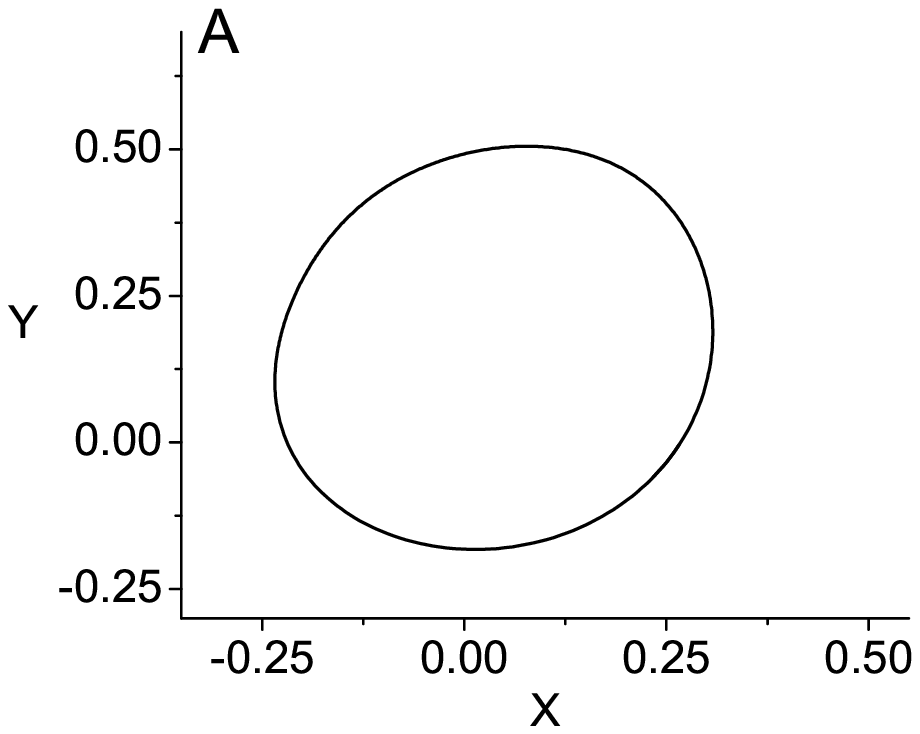}}
\scalebox{0.7}{\includegraphics{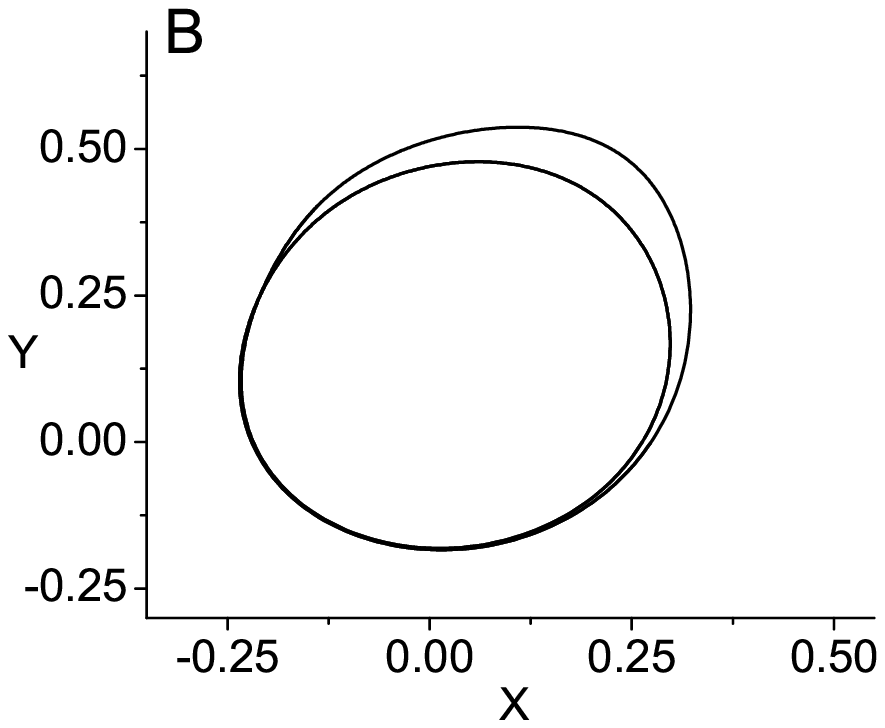}}
\end{center}
\begin{center}
\scalebox{0.7}{\includegraphics{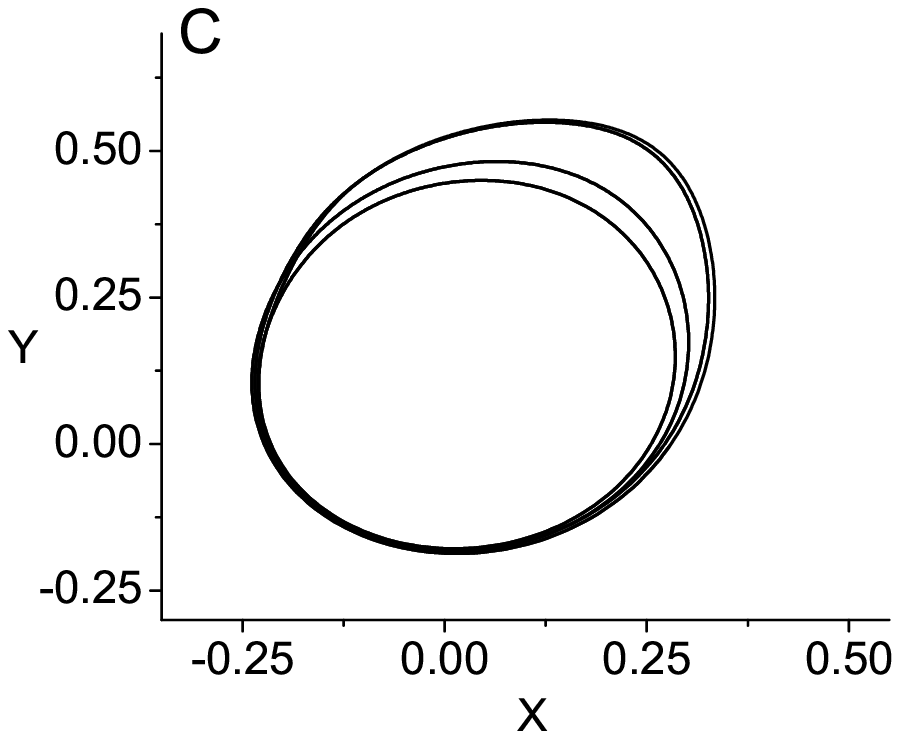}}
\scalebox{0.7}{\includegraphics{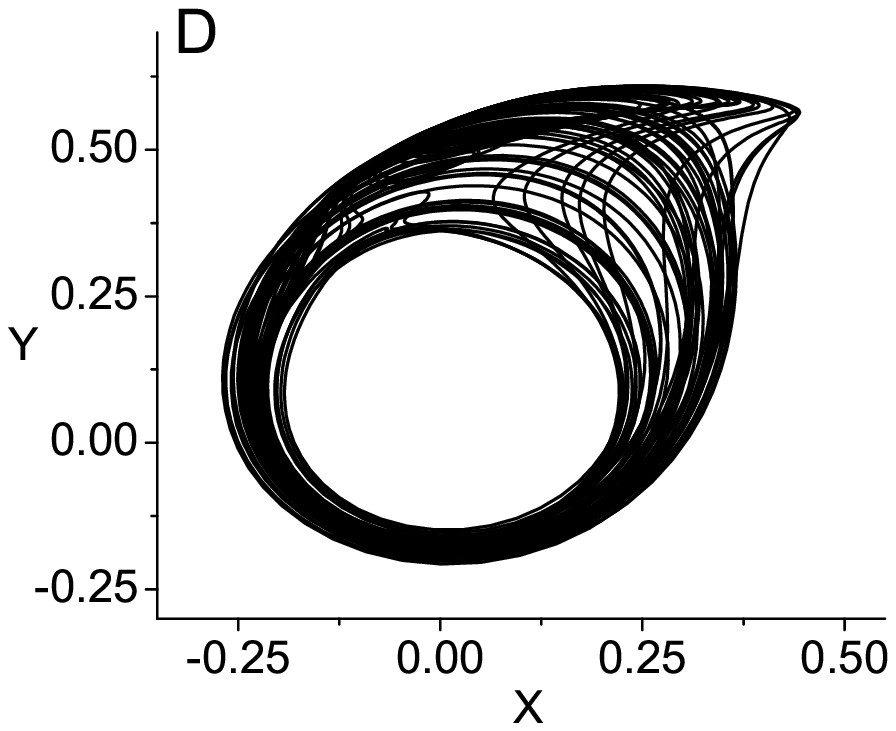}}
\end{center}
\begin{center}
\caption{
Period doubling and chaos in the time-varying network for case (d). The panels show $Y=\rho \sin\psi$ versus $X=\rho \cos \psi$ for A=0.55 (A), 0.56 (B), 0.61 (C), and 0.65 (D), with 
$\omega=1.29, \Delta=0.8$, and $K_0=4$.}
\label{period_double_fig}
\end{center}
\end{figure}

It is easier to visualize this bifurcation sequence using a Poincar\'{e} surface of section with $\psi=\psi_0$ chosen appropriately, so that limit cycles appear as fixed points. We chose trajectory crossings through $\psi_0=1.1$ radians with $\dot{\psi}>0$ and obtained the bifurcation diagram shown in Figure \ref{bifurcation_fig}. The cascade described above is clearly visible in the range $0.41\leq A \leq0.62$. Antimonotonicity is also evident \cite{antimonotonicity}: chaos is destroyed through a sequence of reverse period-doubling bifurcations in the range $0.82 \geq A \geq 0.982$, returning to a full-rotation, 1:1 limit cycle. We have also observed several other limit cycles with different locking ratios created through similar saddle-node bifurcations. These orbits follow their own period-doubling cascades into chaos (e.g., near $A=0.5075$ (1:4) and $A=0.529$ (1:3); note that these are small and are not clearly apparent in the figure). Note also that the small limit cycle (libration) created in the weak time-variation regime, visible on the left near $\rho=0.68$, terminates at a saddle node bifurcation at $A=0.545$.

\begin{figure}
\begin{center}
\scalebox{0.5}{\includegraphics{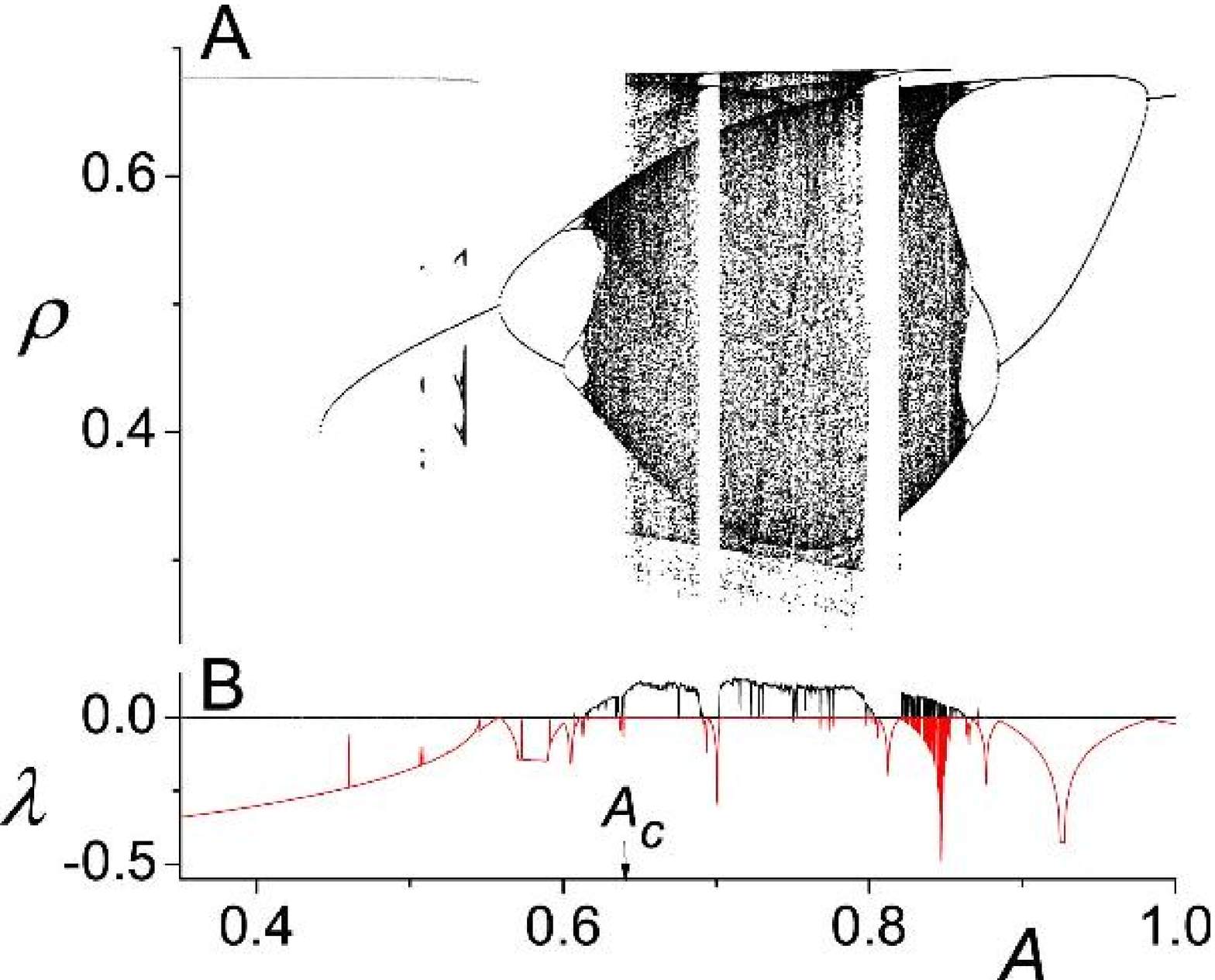}}
\caption{(Color online) A: Bifurcation diagram, obtained using a Poincar\'{e} surface of section, showing the magnitude $\rho$ of the macroscopic mean field versus the amplitude $A$ of coupling variation, in case (d). B: Plot of the two largest Lyapunov exponents versus $A$. Both panels were computed using the reduced system, Eq.~(\ref{bimodal_eq}).}
\label{bifurcation_fig}
\end{center}
\end{figure}

\begin{figure}
\begin{center}
\scalebox{0.7}{\includegraphics{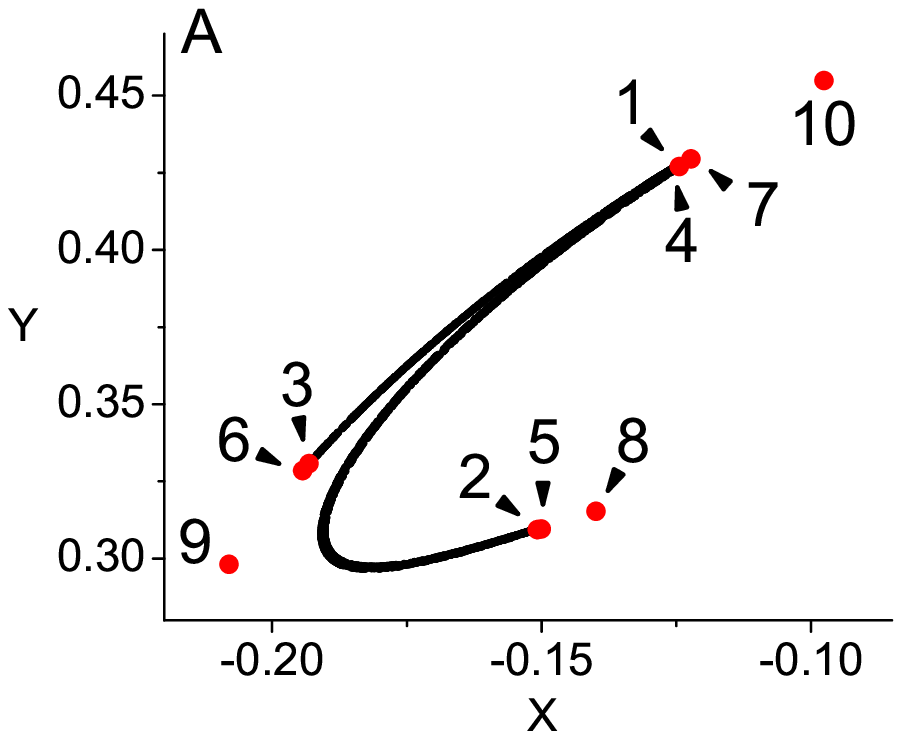}}
\scalebox{0.7}{\includegraphics{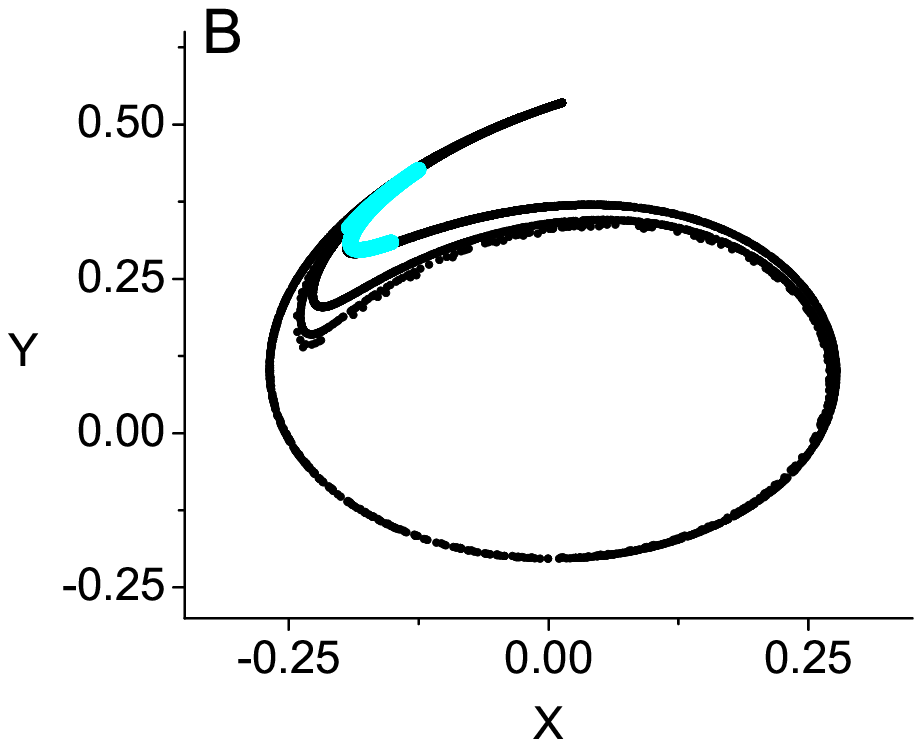}}
\caption{(Color online) a) The chaotic attractor of the macroscopic mean-field slightly above and below the interior crisis value at $A_c=0.641$. For these diagrams, a stroboscopic surface of section, $(\rho((n+1)\tau),\psi((n+1)\tau))$ vs. $(\rho(n\tau),\psi(n\tau))$, was used. A: the attractor for $A \lesssim A_c$. Superimposed on this is a trajectory segment for $A \gtrsim A_c$ showing escape via the mediating period-three orbit. B: the expanded attractor for $A \gtrsim A_c$. The pre-crisis attractor is overlaid in blue.}
\label{crisis_fig}
\end{center}
\end{figure}

The lower panel of Figure \ref{period_double_fig} shows the two largest Lyapunov exponents, calculated using the reduced system (Eq. \ref{bimodal_eq}).
Regions in which one is positive are clearly evident, as are regions in which neither is positive, e.g., in one of the infinitely-many periodic windows \cite{Barretowindows,GS}.

Focusing on the main 1:1 phase-locked bifurcation branch, we see that the chaotic bands merge in the usual way and become a one-piece chaotic attractor near $A=0.627$. Then, at $A_c=0.641$, the chaotic attractor explodes in size through an interior crisis. This occurs when the attractor touches the stable manifold of a coexisting unstable period-three orbit.  Figure \ref{crisis_fig}(A) shows a stroboscopic map in $(\rho,\psi)$ (points sampled every $\tau$ time units) showing the smaller chaotic attractor when $A\lesssim A_c$. For $A\gtrsim A_c$, a typical orbit remains near the core smaller attractor most of the time, but intermittently, bursts occur during which the trajectory visits a more expanded region of state space (see Fig. \ref{crisis_fig} (B)). A sequence of iterates (numbered red dots) during one of the these bursts is shown in panel (a) as the orbit transiently visits the mediating unstable period-three orbit. 
Interestingly, this crisis results in a loss of synchronization between the relative phase difference $\psi$ and the phase of the coupling strength variation.
Before the crisis (Fig. \ref{crisis_fig} (A)), the attractor is restricted to a limited range of $\psi$ in the stroboscopic map (note the axes), indicating that the macroscopic dynamics is phase-locked to the external periodic modulation. After the crisis (Fig. \ref{crisis_fig} (B)), the attractor's phase $\psi$ ranges throughout the entire 0 to 2$\pi$ range. Thus, the macroscopic mean-field slips in phase with respect to the coupling modulation \cite{phaseslips}.

We explicitly demonstrate the existence of a chaotic set in the macroscopic mean field using the stroboscopic map described above. Setting $A=0.65$, we choose the set of initial conditions $(\rho(n\tau),\psi(n\tau))$ indicated by the curved trapezoidal region in Fig. \ref{horseshoe_fig}. The next iterate $(\rho((n+1)\tau),\psi((n+1)\tau))$ of this set under the stroboscopic map (equivalent to integrating these initial conditions forward by $\tau$ time units) forms the elongated and curved region shown in the figure. These intersect in the manner of a Smale horseshoe, thus indicating that the reduced equations for the time-varying network contain a chaotic set \cite{Smale67,Ottbook}.
\begin{figure}
\begin{center}
\scalebox{0.6}{\includegraphics{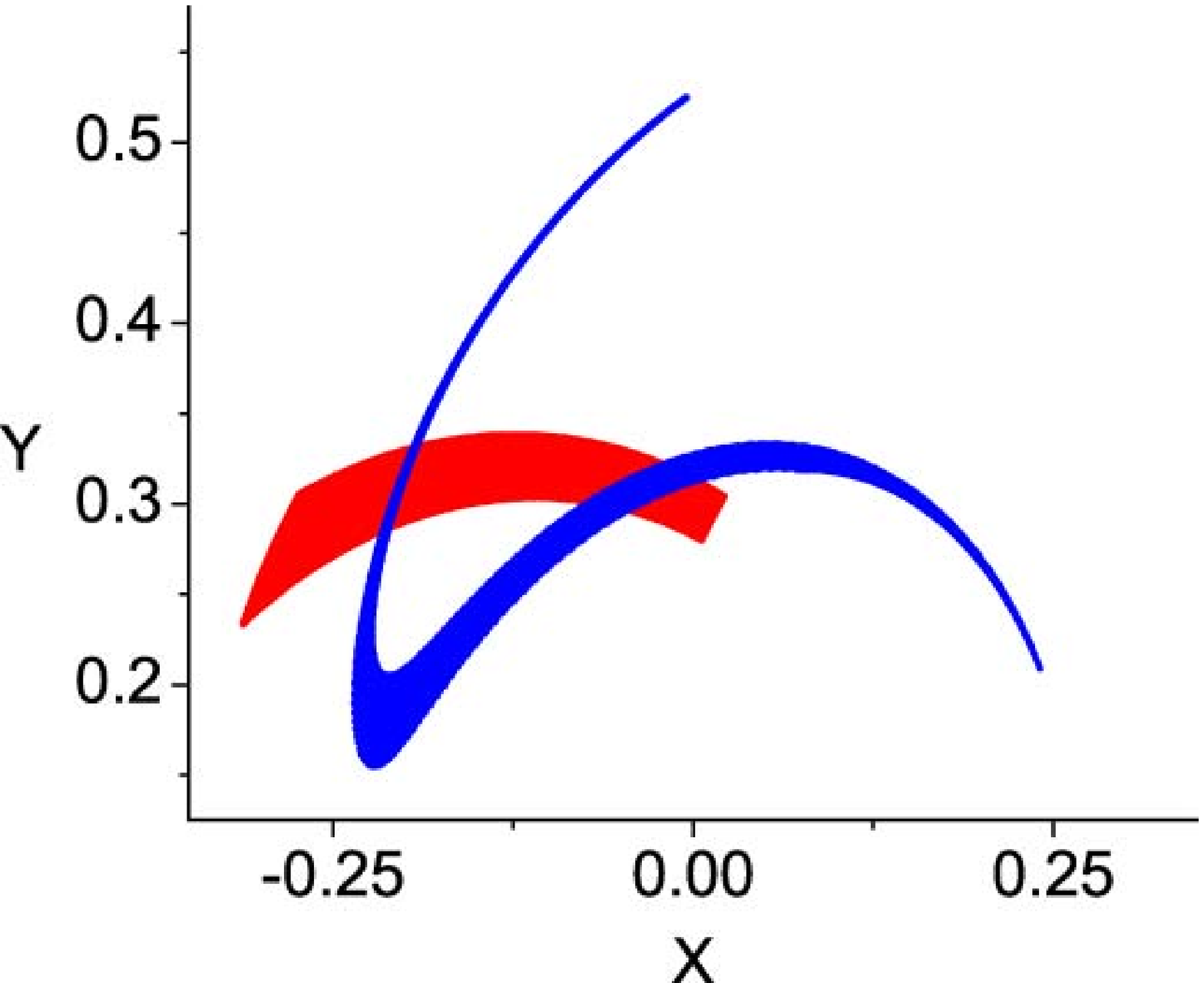}}
\caption{(Color online) A horseshoe in the stroboscopic map of the reduced equation Eq.~(\ref{bimodal_eq}) for $A=0.65$.}
\label{horseshoe_fig}
\end{center}
\end{figure}

All of the above examples had the period $\tau$ of coupling variation set to $5$. Continuing with the parameters for case (d), we fix $A=0.65$ and examine the effect of changing $\tau$. The results appear in Fig.~\ref{taubif_fig}, which shows a bifurcation diagram of $\rho$ versus $\tau$ obtained via Poincar\'{e} surface of section as above (again with $\psi_0=1.1$ radians subject to $\dot{\psi}>0$). For intermediate values of $\tau$, the macroscopic mean-field once again exhibits period-doubling cascades, antimonoticity, crises, and periodic windows. At the extremes, the expected behavior discussed earlier is evident. For slow coupling variation ($6.67<\tau \lessapprox 16.0$), the system exhibits a simple periodic orbit, in which the trajectory follows the pseudo-static attracting equilibrium and limit cycle well; this
appears as the fixed point on the right of Fig.~\ref{taubif_fig}. For very fast coupling variation ($\tau < 4.1$), the fixed point on the left of Fig.~\ref{taubif_fig} corresponds to a very small periodic orbit (libration) that approximates the equilibrium of the static system with $K= \left< K(t) \right>=K_0$. This can be seen in the movies of Figs.~\ref{movie_fast} and \ref{movie_vfs} in the Appendix, in which the static system's equilibrium is marked with an ``X".
\begin{figure}
\begin{center}
\scalebox{0.7}{\includegraphics{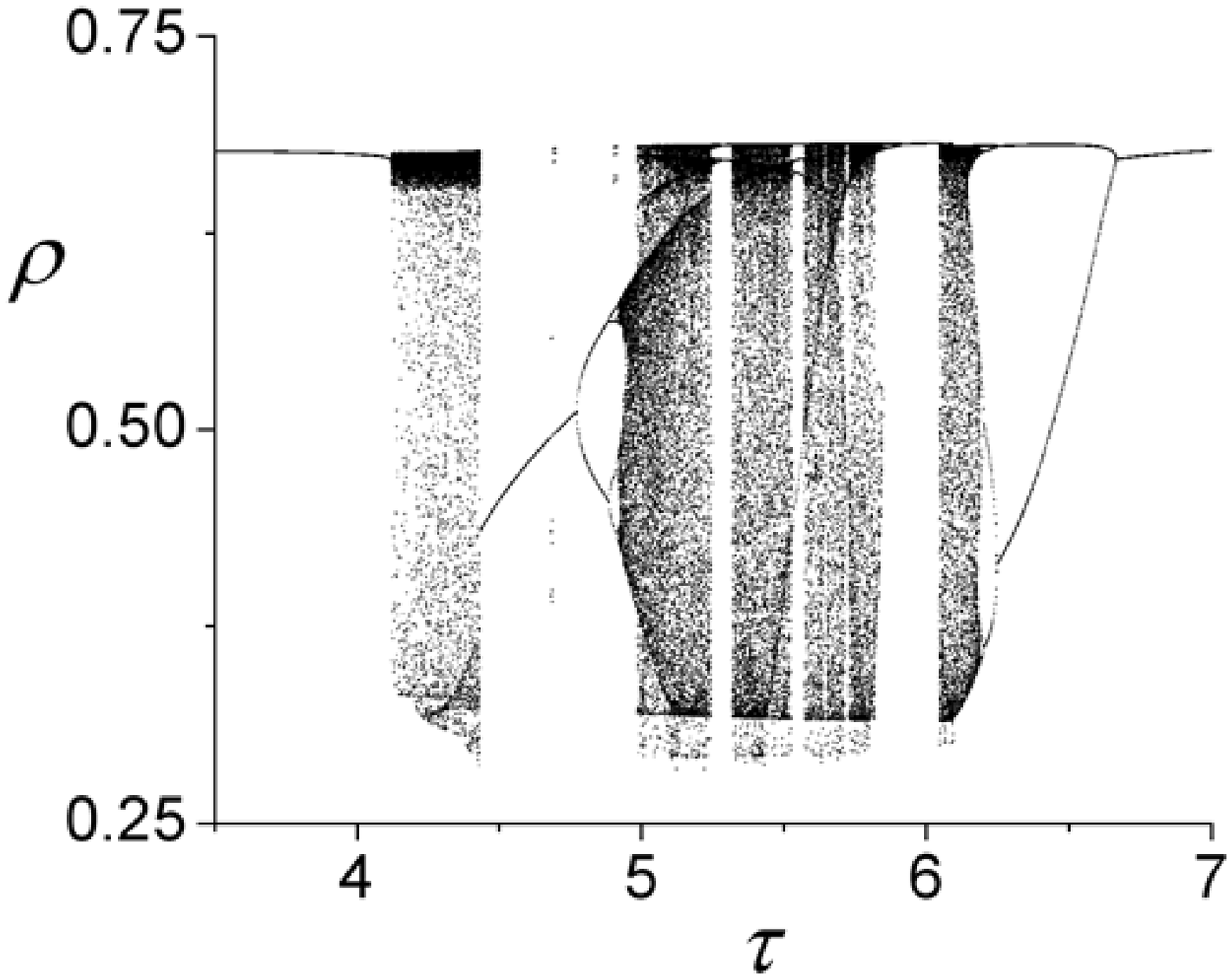}}
\caption{The bifurcation diagram for the magnitude $\rho$ of the macroscopic mean-field versus the period $\tau$ of the coupling time variation. A Poincar\'{e} surface of section is used. $A=0.65$, $\omega_0=0.8$, and $\Delta=1.29$.}
\label{taubif_fig}
\end{center}
\end{figure}

Finally, we briefly consider case (e), in which we return to varying $A$ with $\tau=5$. Here, the parameter sweep 
only crosses the SNIPER bifurcation (see Fig.~\ref{phase_diagram_fig}). Figure \ref{sniperbifdiag} shows that
period-doubling cascades to chaos, crises, and regions with positive Lyapunov exponents are once again evident,
although in a smaller range of $A$.
\begin{figure}
\begin{center}
\scalebox{0.6}{\includegraphics{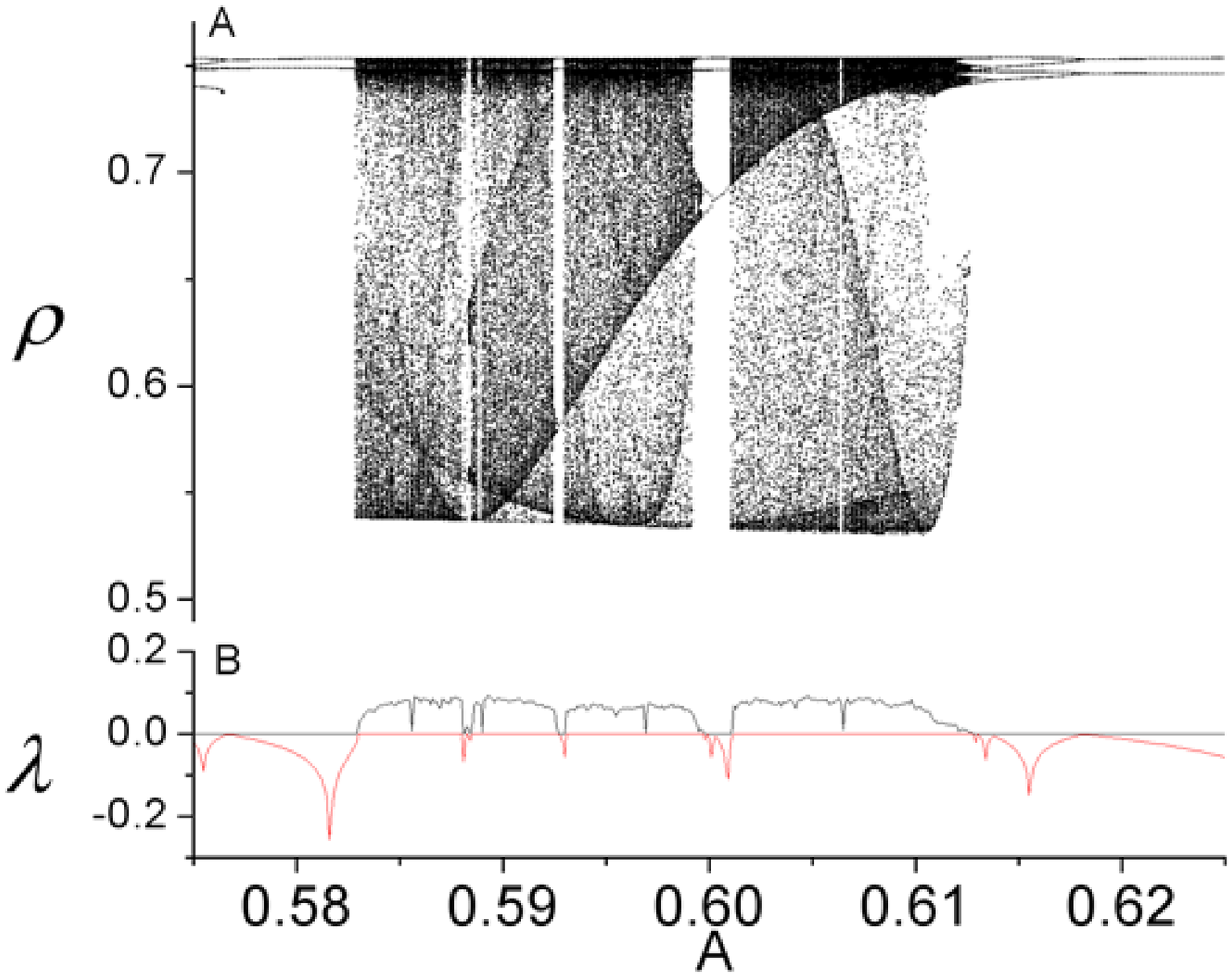}}
\caption{(Color online) Bifurcation diagram and Lyapunov exponents for case (e); compare Fig.\ref{bifurcation_fig}. $\omega_0=1.4$, $\Delta=0.65$.}
\label{sniperbifdiag}
\end{center}
\end{figure}

\section{Discussion}

We have demonstrated that macroscopic chaos occurs in the thermodynamic limit of the bimodal Kuramoto system with periodic time-variation of the coupling strength. This is interesting because it is known that an infinite but autonomous network of phase oscillators with a symmetric bimodal natural frequency distribution cannot exhibit chaos. This follows from analysis based on the OA reduction procedure, which shows that the latter system is effectively two-dimensional at the level of the mean field dynamics. Thus, chaos is impossible. However, by introducing a periodic time-variation of the coupling strength, a third dimension is added to the system, and under proper circumstances, chaos is found. 

To clarify the circumstances under which macroscopic chaos is possible in the time-varying system, we have developed the concept of pseudo-static attractors that act as ``moving targets'' which influence the trajectory of the macroscopic mean field. These pseudo-static attractors are the attractors of the static system for any given fixed coupling strength $K$. As $K$ varies in time, these become pseudo-attractors that move about in state space depending on the frequency (or period) of variation. As a result, the trajectory of the time-varying system may or may not be able to ``keep up" with the motion of the pseudo-attractors. We find that interesting complicated dynamics arise when the frequency of variation is such that the trajectory of the macroscopic mean field is essentially frustrated in that it is not able to ever reach the moving pseudo-attractors. In this situation, the trajectory is dominated by what we call transitory excursions. The frequencies for which complicated dynamics arise should, in some sense, be commensurate with the rates of attraction to the pseudo-static attractors. In the examples that we have described here, we observed period-doubling cascades to chaos, crises, horseshoes, and other complex nonlinear dynamical features.

Our understanding of this mechanism is based on detailed knowledge of the asymptotic structures and bifurcations of the macroscopic mean field present in the static bimodal Kuramoto system, published earlier \cite{Martens09}. Our observations suggest the following conjecture. In order to generate macroscopic chaos on the antsatz manifold, the static system must exhibit the following dynamical features:
(1) {\bf Topological Choice}: Within the range of parametric variation (in our case, $K_0 \pm A$), the static system must possess at least two attracting macroscopic structures that are separated in state space. The separation is necessary in order to \emph{permit} substantial transitory excursions. In our examples (case (d) and (e)), the static system has an attracting equilibrium and/or an attracting limit cycle.
(2) {\bf Switching}: Within the range of parameter variation, there must exist bifurcations of the static system that create and destroy these attracting macroscopic structures. In our examples, the attracting equilibrium is created/destroyed through a saddle-node bifurcation, and the attracting limit cycle is created/destroyed through a homoclinic bifurcation. Furthermore, the parametric variation should include a range in which these macroscopic attractors exist independently. This is necessary in order to \emph{cause} transitory excursions. For case (d) (see Fig. \ref{sweepbifdiag} (A)), the equilibrium is the only attractor in the static system for $K > 3.953$, and the limit cycle is the only attractor in the static system for $K \in [3.2,3.935]$. Although both of these attractors coexist in a very small interval for case (d), 
there is no such coexistence in case (e), demonstrating that multistability is not necessary for complex behavior to arise in the time-varying system (Fig.~\ref{sniperbifdiag}).

Finally, we note that this general mechanism need not be restricted to non-autonomous systems. We suspect that a similar mechanism could arise in any system that has a dynamic interplay between fast and slow dynamics under the conditions described above. Indeed, we have found complicated behavior, including chaos, in an autonomous model of a neuron featuring fast spiking dynamics which are modulated by slow autonomous ion concentration dynamics \cite{BarretoCressman}. We leave this for future investigation.

\section{Appendix}

The following movies, available online, help to visualize the dynamics described in the main text. In all movies, case (d) is considered ($\omega_0=0.8$, $\Delta=1.29$) with $A=0.65$. The pseudo-static attractors are shown in black (compare Fig.~\ref{sweepbifdiag}), and the trajectory of the time-varying system is shown in cyan, being traced by the red dot.

Figure \ref{movie_slow} shows an extreme case of slow variation ($\tau=50$), and is reminiscent of a bursting neuron.
\begin{figure}
\begin{center}
\scalebox{0.8}{\includegraphics{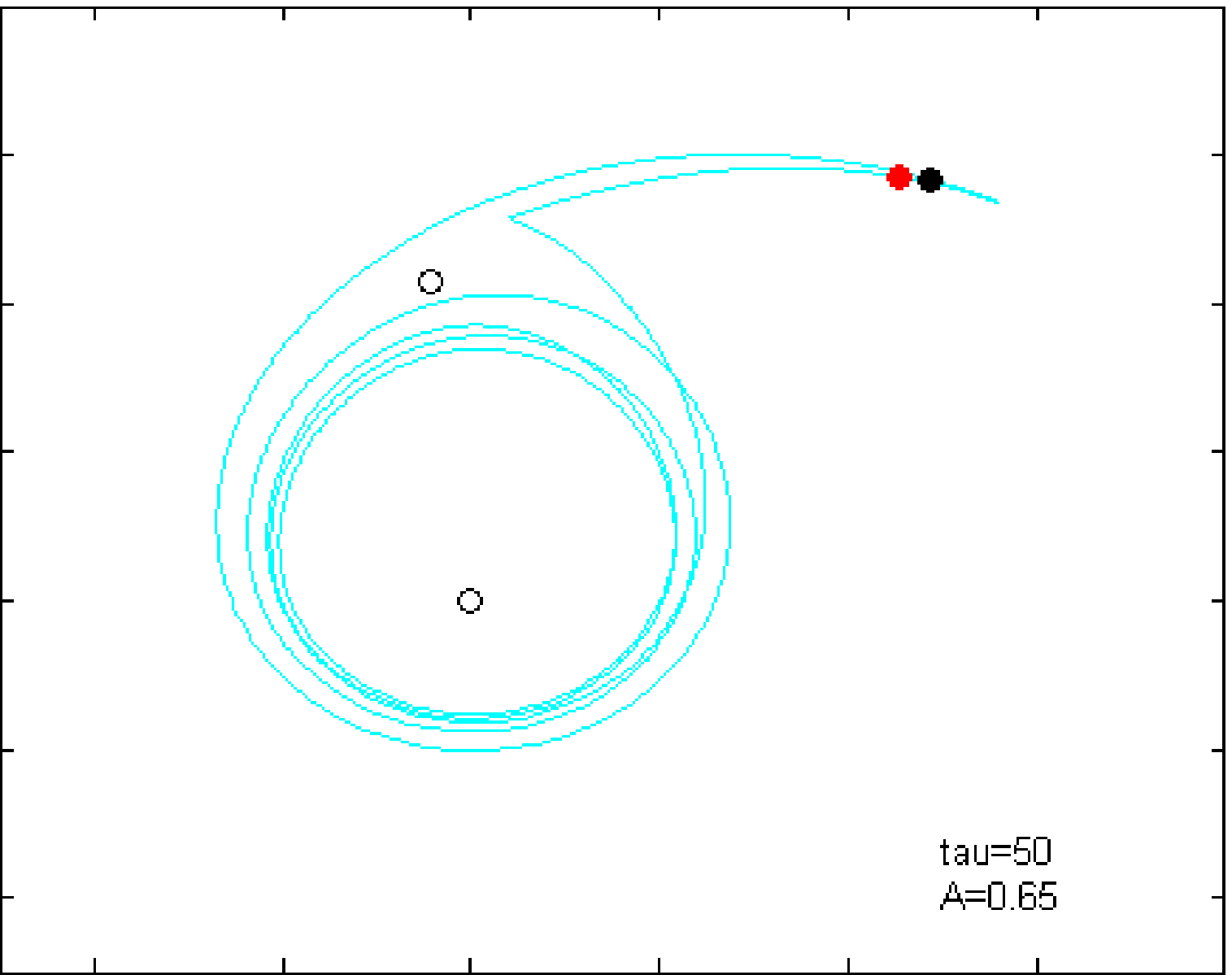}}
\caption{Very slow parameter variation with $\tau=50.0$ (enhanced online).}
\label{movie_slow}
\end{center}
\end{figure}
Figures \ref{movie_fast} and \ref{movie_vfs} show the fast ($\tau=1$) and very fast ($\tau=0.1$) variation case, illustrating the time-varying system's approach to the equilibrium of the static system with $K=\left< K(t) \right>=K_0$. The location of the latter is marked with a black ``X".
\begin{figure}
\begin{center}
\scalebox{0.8}{\includegraphics{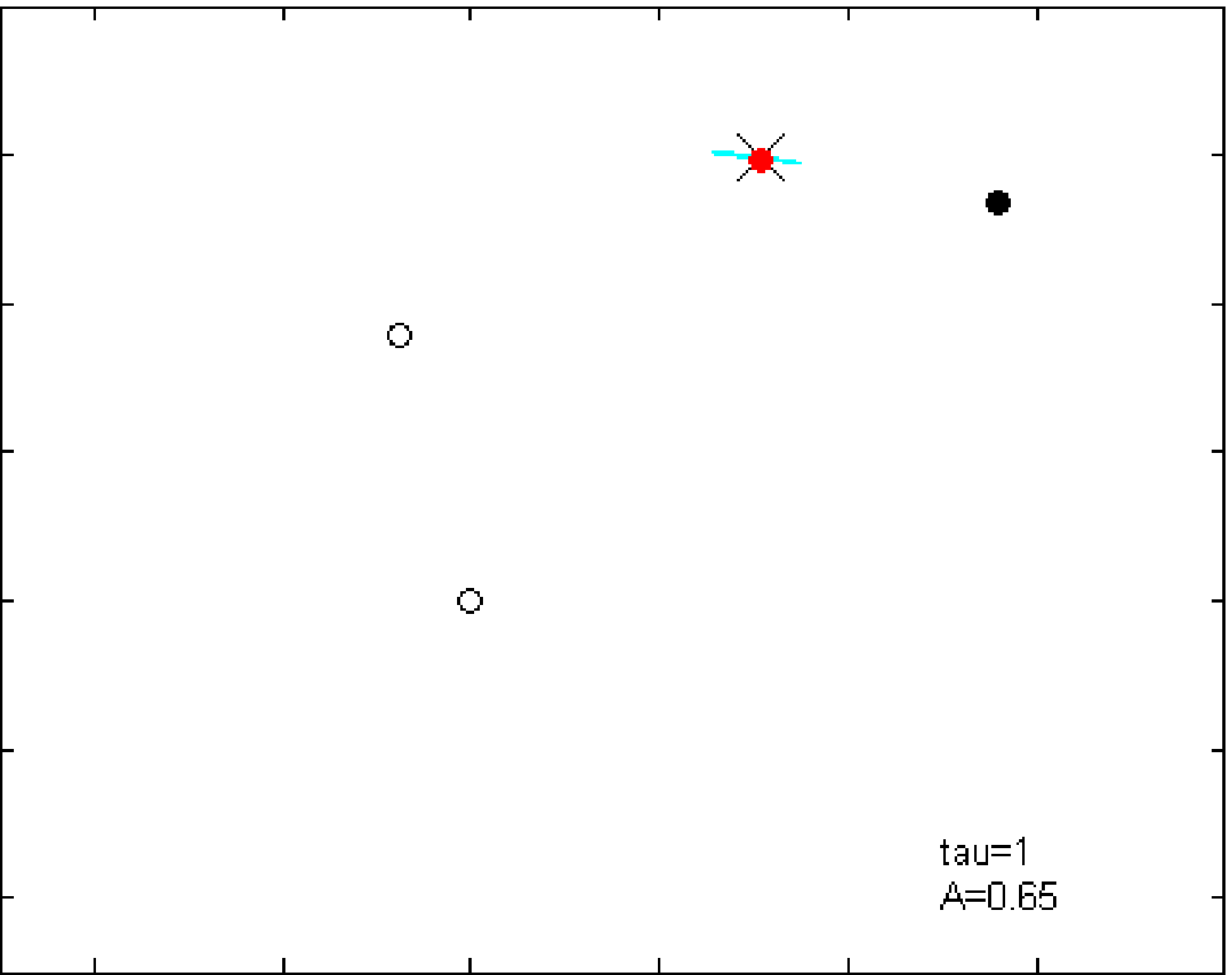}}
\caption{Fast parameter variation with $\tau=1.0$ (enhanced online).}
\label{movie_fast}
\end{center}
\end{figure}
\begin{figure}
\begin{center}
\scalebox{0.8}{\includegraphics{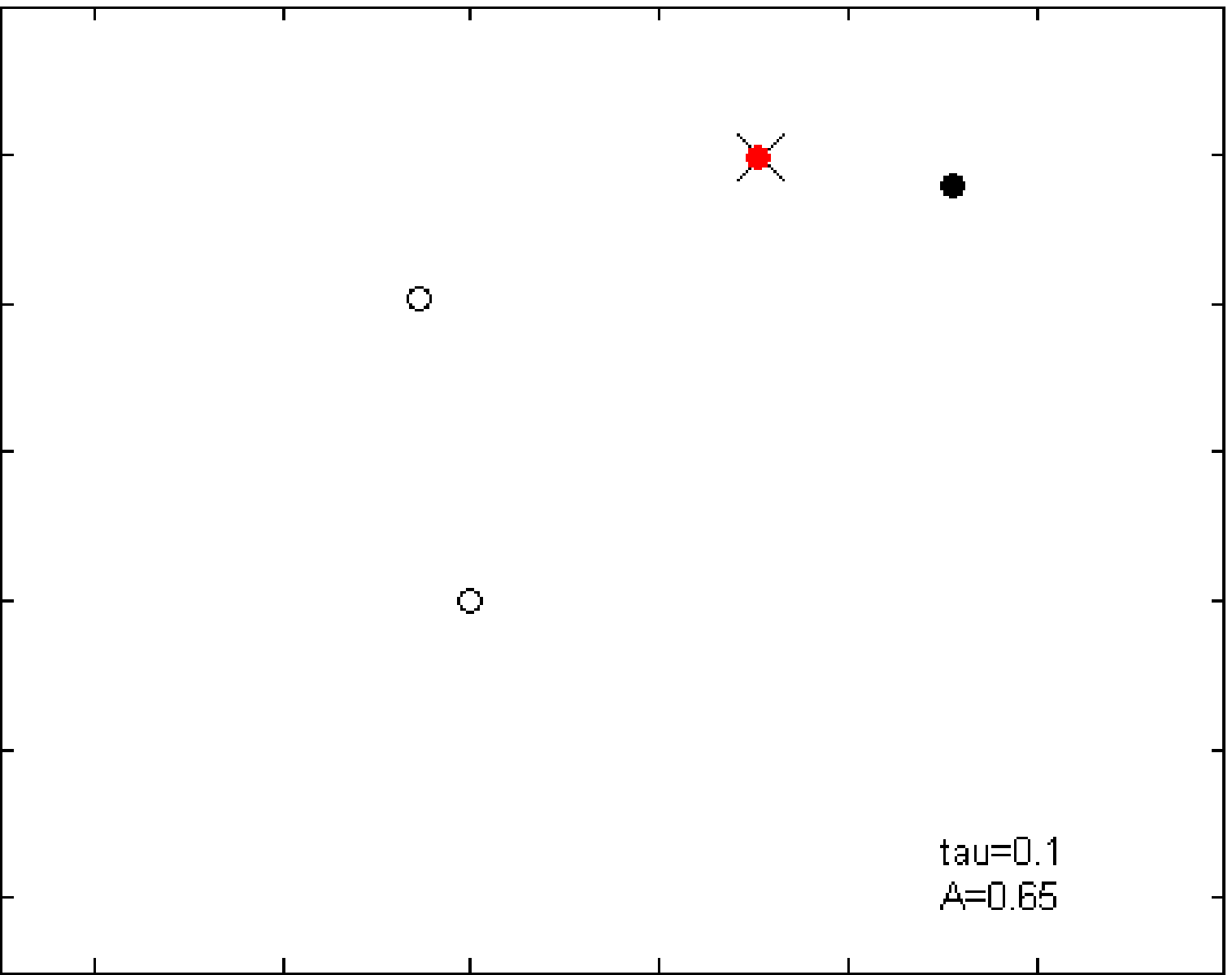}}
\caption{Very fast parameter variation with $\tau=0.1$ (enhanced online).}
\label{movie_vfs}
\end{center}
\end{figure}
Finally, the chaotic trajectory obtained for $\tau=5$ is shown in Fig.~\ref{movie_chaos}.
\begin{figure}
\begin{center}
\scalebox{0.8}{\includegraphics{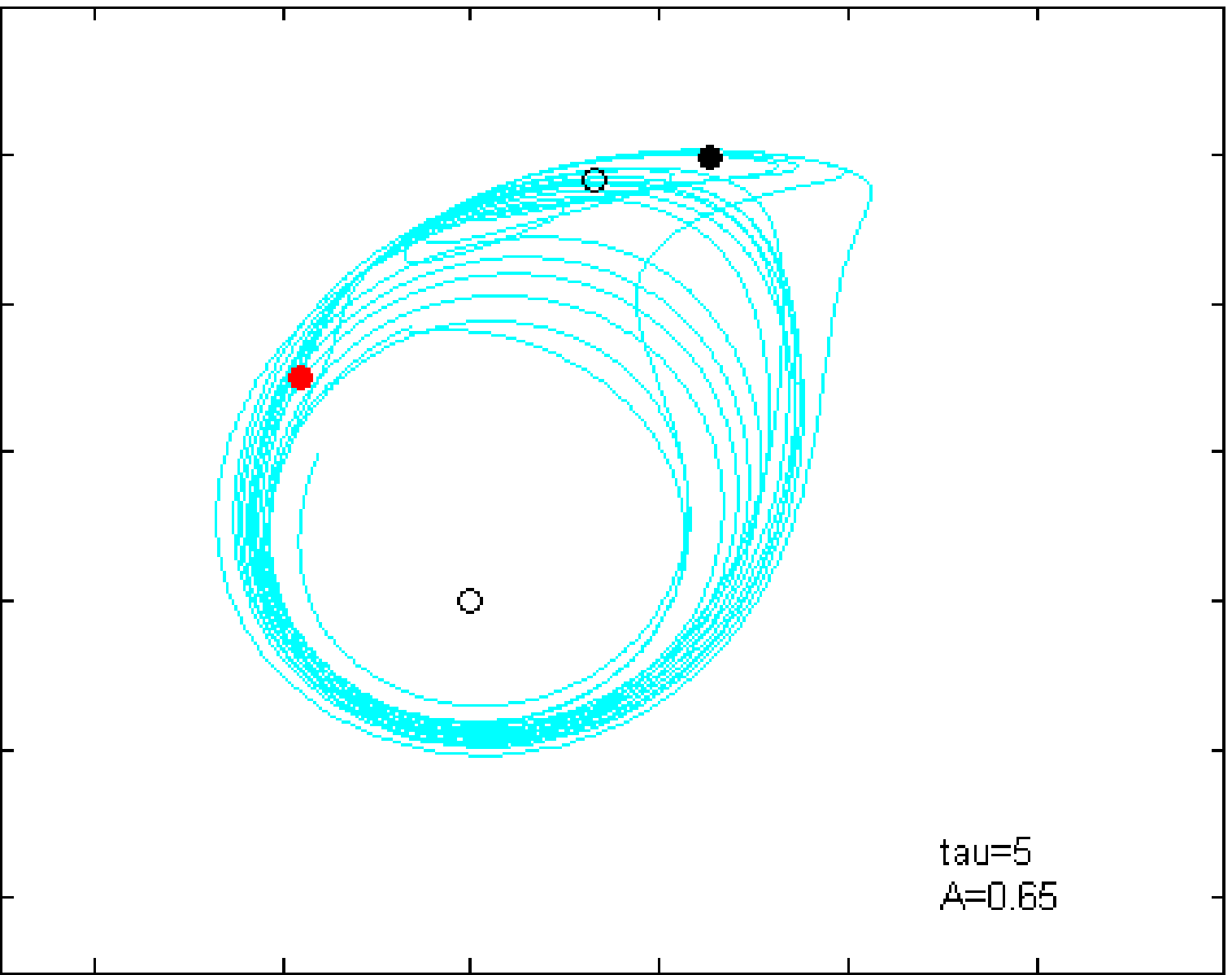}}
\caption{Chaos: intermediate parameter variation with $\tau=5.0$ (enhanced online).}
\label{movie_chaos}
\end{center}
\end{figure}


\begin{thebibliography}{}

\bibitem{generalsynch}
A.T. Winfree, {\it The Geometry of Biological Time}, (Springer, New York, 1980);
S. Yamaguchi, {\it et al.}, Science {\bf 302}, 1408 (2003);
I.Z. Kiss, Y. Zhai, and J.L. Hudson, Science {\bf 296}, 1676 (2002);
J. Pantaleone, Phys. Rev. D {\bf 58}, 073002 (1998);
K. Wiesenfeld and J.W. Swift, Phys. Rev. E {\bf 51}, 1020 (1995).

\bibitem{Kuramoto75} Y. Kuramoto, in {\it International Symposium on Mathematical Problems in Theoretical Physics}, Lecture Notes in Physics, Vol. 39, edited by H. Araki (Springer-Verlag, Berlin, 1975).

\bibitem{Kuramoto84}
Y. Kuramoto, {\it Chemical Oscillations, Waves and Turbulence} (Springer, 1984), pp. 75--76.

\bibitem{Crawford94}
J.D. Crawford, J. Stat. Phys. {\bf 74}, 1047 (1994).

\bibitem{Strogatz00}
S.H. Strogatz, Physica D {\bf 143}, 1 (2000).

\bibitem{Acebron05}
J.A. Acebr\'{o}n, L.L. Bonilla, C.J. P\'{e}rez Vicente, F. Ritort, and R. Spigler, Rev. Mod. Phys. {\bf 77}, 137 (2005).


\bibitem{collectivechaos}
A.S. Pikovsky, M.G. Rosenblum, J. Kurths, Europhys. Lett. {\bf 34}, 165 (1996);
H. Sakaguchi, Phys. Rev. E {\bf 61}, 7212 (2000);
E. Ott, P. So, E. Barreto, and T. Antonsen, Physica D {\bf 173}, 29-51 (2002);

\bibitem{chaoticmap}
A.S. Pikovsky and J. Kurths, Phys. Rev. Letts. {\bf 72}, 1644 (1994);
T. Shibata and K. Kaneko, Phys. Rev. Letts. {\bf 81}, 4116 (1998);
T. Shibata, T. Chawanya, and K. Kaneko, Phys. Rev. Letts. {\bf 82}, 4424 (1999);
M. Cencini, M. Falcioni, D. Vergni, and V. Vulpiani, Physica D {\bf 130}, 58 (1999);
I. Omelchenko, Y. Maistrenko, P. H\"ovel, and E. Sch\"oll, Phys. Rev. Lett. {\bf 106} 234102 (2011).

\bibitem{OAreduction08}
E. Ott and T.M. Antonsen, CHAOS {\bf 18}, 037113 (2008).

\bibitem{OAreduction09}
E. Ott and T.M. Antonsen, CHAOS {\bf 19}, 023117 (2009).

\bibitem{Martens09}
E.A. Martens, E. Barreto, S.H. Strogatz, E. Ott, P. So, and T.M. Antonsen, Phys. Rev. E {\bf 79}, 026204 (2009).

\bibitem{Marvel09}
S.A. Marvel and S.H. Strogatz, CHAOS {\bf 19}, 013132 (2009).

\bibitem{Pikovsky08}
A. Pikovsky and M. Rosenblum, Phys. Rev. Lett. {\bf 101}, 264103 (2008).

\bibitem{Paz06}
E. Montbri\'o, D. Paz\'o, and J. Schmidt, Phys. Rev. E {\bf 74}, 056201 (2006).

\bibitem{PazoBimodal}
D. Paz\'o and E. Montbri\'o, Phys. Rev. E {\bf 80}, 046215 (2009).

\bibitem{oaothers}
D.M. Abrams, R. Mirollo, S.H. Strogatz, D.A. Wiley, Phys. Rev. Letts {\bf 101}, 084103 (2008);
T.M. Antonsen, Jr. R.T. Faghih, M. Girvan, E. Ott, and J. Platig, CHAOS {\bf 18}, 037113 (2008);
L.F. Lafuerza, P. Colet, R. Toral, Phys. Rev. Letts {\bf 105}, 084101 (2010);
A. Pikovsky and M. Rosenblum, Phys. Rev. Letts. {\bf 101}, 264103 (2008);
W.S. Lee, E. Ott, and T.M. Antonsen, Phys. Rev. Letts. {\bf 103}, 044101 (2009);
L.M. Alonso, J.A. Alliende, and G.B. Mindlin, Eur. Phys. J. D {\bf 60}, 361 (2010);
C. R. Laing, Chaos {\bf 19}, 013113, (2009).
C. R. Laing, Physica D {\bf 238}, 1569-1588 (2009).


\bibitem{Kuramoto93}
N. Nakagawa and Y. Kuramoto, Prog. Theo. Phys {\bf 89}, 313 (1993).

\bibitem{Kuramoto94}
N. Nakagawa and Y. Kuramoto, Physica D {\bf 75}, 74 (1994).

\bibitem{Matthews91}
P.C. Matthews, R.E. Mirollo, and S.H. Strogatz, Physica D {\bf 52}, 293 (1991).

\bibitem{Takeuchi09}
K.A. Takeuchi, F. Ginelli, and H. Chat'{e}, Phys. Rev. Letts. {\bf 103}, 154103 (2009).

\bibitem{Popovych05}
Y.L. Maistrenko, O.V. Popovych, and P.A. Tass, Int. J. Bifurcation and Chaos {\bf 15}, 3457 (2005).

\bibitem{Rapisarda09}
G. Miritello, A. Pluchino, and A. Rapisarda, EPL {\bf 85}, 10007 (2009).

\bibitem{Golomb92}
D. Golomb, D. Hansel, B. Shraiman, and H. Sompolinsky, Phys. Rev. A {\bf 45}, 3316 (1992).

\bibitem{Watanabe94}
S. Watanabe and S.H. Strogatz, Physica D {\bf 74}, 197 (1994).

\bibitem{Bonilla92}
L.L. Bonilla, J.C. Neu, and R. Spigler, J. Stat. Phys {\bf 67}, 313 (1992).

\bibitem{So08}
P. So, B.C. Cotton, and E. Barreto, CHAOS {\bf 18}, 037114 (2008).

\bibitem{Cotton10}
E. Barreto, B.C. Cotton, P. So (in preparation). 

\bibitem{noteattract}
One might ask is whether the dynamics for the macroscopic mean field described by Eqs. (\ref{solution_eq}) and (\ref{solution_mean_eq}) truly represent the long-time dynamics for the full network since the ansatz Eq. (\ref{ansatz_eq}) only represents a small subspace within the space of all possible density functions.  Ott and Antonsen (Ref \cite{OAreduction09}) showed that under some rather generic conditions, the submanifold defined by the ansatz is indeed attracting as long as the distribution of the natural frequencies for the phase oscillators has a nonzero spread. In particular, the macroscopic mean field $z(t)$ can be shown to exponentially approach the quantity given by $\int_{-\infty}^{\infty}\alpha^*(\omega,t)g(\omega)d\omega$ at large $t$ and $\alpha(\omega,t)$ will evolves according to Eq. (\ref{solution_eq}). 

\bibitem{Barreto07}
E. Barreto, B. Hunt, E. Ott and P. So, Phys. Rev. E {\bf 77}, 036107 (2008).

\bibitem{Gluckenheimer_holmes}
J. Guckenheimer and P. Holmes, {\it Nonlinear Oscillations, Dynamical Systems, and Bifurcations of Vector Fields}, (Springer-Verlag, 1983), pp. 166--180.

\bibitem{Sander_verhulst}
J.A. Sanders and F. Verhulst, {\it Averaging Methods in Nonlinear Dynamical Systems}, 2nd Ed., (Springer, 2007), pp. 18--19.

\bibitem{trapping}
One can see from the $\rho$ equation in Eq. (\ref{bimodal_eq}) that $\dot{\rho}<0$ at $\rho=1$ for all choices of the parameters including the time varying situation.  When the incoherent state becomes an unstable focus after passing through the Hopf bifurcation, the annulus region defined by $0<\rho<1$ becomes a trapping region for the network dynamics.  For the time static network ($A=0$), the asymptotic state is a periodic orbit and for the time-varying network with two incommensurate intrinsic frequencies, the macroscopic state tends toward a quasi-periodic state within this annulus region.

\bibitem{quasiperiodic_future} Analysis of the phase locking and the Arnold tongue structure for this time-varying bimodal Kuramoto network will be reported elsewhere.

\bibitem{Verhulst96}
F. Verhulst, {\it Nonlinear Differential Equations and Dynamical Systems}, 2nd Ed., (Springer, 1996), pp. 136--162.

\bibitem{fastswitching}
I.V. Belykh, V.N. Belykh, and M. Hasler, Physica D {\bf 195}, 188 (2004);
J.D. Skufca and E.M. Bollt, Math. Biosci. {\bf 1} 347 (2004);

\bibitem{Feigenbaum}
M.J. Feigenbaum, J. Stat. Phys. {\bf 19}, 25 (1978).

\bibitem{Ottbook}
E. Ott, {\it Chaos in Dynamical Systems}, 2nd Ed., (Cambridge University Press, 1993), pp. 32--45.

\bibitem{antimonotonicity}
I. Kan, H. Kocak, and J.A. Yorke, Ann. of Math, {\bf 136}, 219 (1992);
S.P. Dawson, C. Grebogi, and H. Kocak, {\bf 48}, 1676 (1993).


\bibitem{Barretowindows}
E. Barreto, B. Hunt, C. Grebogi, and J.A. Yorke, Phys. Rev. Lett. {\bf 78}, 4561--4564 (1997).

\bibitem{GS}
J. Graczyk and G. \'Swi{\c a}tek, Ann. of Math. {\bf 146}, 1--52 (1997).

\bibitem{phaseslips}
M.A. Zaks, E-H Park, M.G. Rosenblum, and J. Kurths,
Phys. Rev. Lett. {\bf 82}, 4228 (1999).

\bibitem{Smale67}
S. Smale, Bulletin of the American Mathematical Society {\bf 73}, 747 (1967).

\bibitem{BarretoCressman}
E. Barreto and J.R. Cressman, Journal of Biological Physics 37, 361-373 (2011).

\end{thebibliography}
\end{document}